\documentclass[a4paper, 12pt, oneside]{article}

\usepackage{aapreprint}

\usepackage{graphicx,wrapfig,float,slashed,subcaption,bbold,bm}
\usepackage{amsmath,amssymb,epsfig,graphicx,xcolor}
\usepackage{hyperref}
\usepackage{epstopdf}
\usepackage{booktabs}
\usepackage{ragged2e}
\usepackage{mciteplus}
\usepackage{amssymb}
\usepackage{diagbox}
\usepackage{float}
\usepackage{tikz} 
\usetikzlibrary{shapes,arrows,positioning,automata,backgrounds,calc,er,patterns}
\usepackage{tikz-feynman}
\tikzfeynmanset{compat=1.0.0}

\newcommand{\nc}{\newcommand}
\nc{\non}{\nonumber}
\nc{\hc}{\hbox {H.c.}}
\nc{\noi}{\noindent}
\nc{\barx}{\bar{x}}
\nc{\pbarn}{\;\hbox {pb}}
\nc{\fbarn}{\;\hbox {fb}}

\nc{\hsp}{\hspace{0.5cm}}
\nc{\lsp}{\hspace{1cm}}
\nc{\Lsp}{\hspace{2cm}}
\nc{\LLsp}{\lsp\lsp}
\nc{\lra}{\longrightarrow}
\nc{\p}{\prime}
\nc{\sgn}{\text{sgn}}
\nc{\ph}{\varphi}
\nc{\op}{{\cal O}}

\newcommand{\vect}[1]{\boldsymbol{#1}}
\nc{\beq}{\begin{equation}}  \nc{\eeq}{\end{equation}}
\nc{\bea}{\begin{eqnarray}}  \nc{\eea}{\end{eqnarray}}
\nc{\baa}{\begin{array}}     \nc{\eaa}{\end{array}}
\nc{\bit}{\begin{itemize}}   \nc{\eit}{\end{itemize}}
\nc{\ben}{\begin{enumerate}} \nc{\een}{\end{enumerate}}
\nc{\bce}{\begin{center}}    \nc{\ece}{\end{center}}
\nc{\bpm}{\begin{pmatrix}}   \nc{\epm}{\end{pmatrix}}
\nc{\bvt}{\begin{verbatim}}  \nc{\evt}{\end{verbatim}}

\def\lsim{\mathrel{\raise.3ex\hbox{$<$\kern-.75em\lower1ex\hbox{$\sim$}}}}
\def\gsim{\mathrel{\raise.3ex\hbox{$>$\kern-.75em\lower1ex\hbox{$\sim$}}}}

\def\udots{\mathinner{\mkern1mu\raise1pt\vbox{\kern7pt\hbox{.}}\mkern2mu\raise4pt\hbox{.}\mkern2mu\raise7pt\hbox{.}\mkern1mu}}

\allowdisplaybreaks

\newcommand\fverb{\setbox\fverbbox=\hbox\bgroup\verb}
\newcommand\fverbdo{\egroup\medskip\noindent%
			\fbox{\unhbox\fverbbox}\ }
\newcommand\fverbit{\egroup\item[\fbox{\unhbox\fverbbox}]}
\newbox\fverbbox

\preprint{\begin{flushright}
UTTG 08-2022\\

\end{flushright}}

\title{Applying Machine Learning Techniques To Intermediate-Length Cascade Decays} 
\author[a]{Maaz Ul Haq,}
\author[a]{Can Kilic,}
\author[a]{Benjamin Lawrence-Sanderson,}
\author[a]{and Ram Purandhar Reddy Sudha}
\affiliation[a]{Theory Group, Department of Physics\\ University of Texas at Austin,
Austin, TX 78712, USA}
\emailAdd{}

\abstract{In the collider phenomenology of extensions of the Standard Model with partner particles, cascade decays occur generically, and they can be challenging to discover when the spectrum of new particles is compressed and the signal cross section is low. Achieving discovery-level significance and measuring the properties of the new particles appearing as intermediate states in the cascade decays is a longstanding problem, with analysis techniques for some decay topologies already optimized. We focus our attention on a benchmark decay topology with four final state particles where there is room for improvement, and where multidimensional analysis techniques have been shown to be effective in the past. We apply Machine Learning techniques in order to identify effective human-level kinematic observables for discovery, spin determination and mass measurement. We quantify the performance of these analyses as a function of the signal size. In agreement with past work, we confirm that the kinematic observable $\Delta_4$ is highly effective.}

\begin{document}
\maketitle
\flushbottom

\section{Introduction}
\label{sec:intro}

While the Standard Model (SM) of particle physics is extremely successful in describing the known particles and their interactions, it is also known to be an incomplete description of fundamental physics. Some of the best studied extensions of the Standard Model that aim to stabilize the electroweak scale or explain the observed dark matter (DM) relic abundance hint at the existence of new degrees of freedom at roughly the TeV scale. Unfortunately, collider searches for such new particles have yielded only null results until now. Therefore, as the Large Hadron Collider (LHC) is getting ready to start its third run, even if new physics is finally discovered, the signal will in all likelihood either have low statistics, or be difficult to distinguish from backgrounds, or possibly both. This makes it all the more crucial that LHC searches be optimized for maximal efficiency with these challenges in mind. Similarly, post-discovery, the measurement of the properties of the new particles, such as their masses, will be challenging for the same reasons.

Among possible final states for new physics at the LHC, supersymmetry (SUSY)-like production and decay channels of color-neutral particles, especially with a compressed spectrum for the new particles, offer good examples for the type of signatures mentioned above, as electroweak-only charged particles have low production cross sections, and compressed spectra result in soft momenta in the final state, so backgrounds cannot be reduced by using hard cuts. Our definition of a SUSY-like channel is that new particles can only be produced in pairs due to a $Z_2$ symmetry, and that each one decays to a lighter new particle plus SM particles, until the lightest new particle is reached, which is collider-stable, a DM candidate and which cannot be detected. Note that this definition applies equally well to scenarios where the new particles have the same spin as their SM partners, such as in the case of extra dimensional models and others. Since there is an invisible particle at the end of any decay chain in such final states, no resonance can be reconstructed from any subset of visible particles. 

For short decay chains, very little kinematic information is available event-by-event. Observables such as missing transverse energy (MET), and transverse mass variables such as $m_T$ and $m_{T2}$~\cite{Lester:1999tx, Allanach:2000kt, Cho:2007dh, Barr:2007hy, Barr:2008ba, Burns:2008va, Barr:2009jv,  Burns:2009zi, Cho:2009zza} fully use this available information and provide the best chance for discovery, and for the mass measurement of the unknown particles. As a result, all available information can be extracted from one-dimensional distributions of a small number of kinematic variables. In the other extreme, for sufficiently long decay chains, there is sufficient kinematic information available in the events for the determination of the complete spectrum by algebraic methods~\cite{Hinchliffe:1998ys, Nojiri:2003tu, Kawagoe:2004rz, Gjelsten:2006as, Cheng:2007xv, Nojiri:2008ir, Cheng:2008mg, Webber:2009vm, Cheng:2009fw, Matchev:2009iw, Autermann:2009js, Kim:2009si, Han:2009ss, Kang:2010mhc, Nojiri:2010dk, Kang:2010hb, Hubisz:2010ur, Cheng:2010yy, Gripaios:2011jm, Han:2012nm, Han:2012nr}. Both of these possibilities have been well studied, and there is little room for improvement. For an extensive review of kinematic variables used in  collider phenomenology, we direct the reader to~\cite{Franceschini:2022vck}.

\begin{figure}
        \centering
        \includegraphics[width=0.5\textwidth]{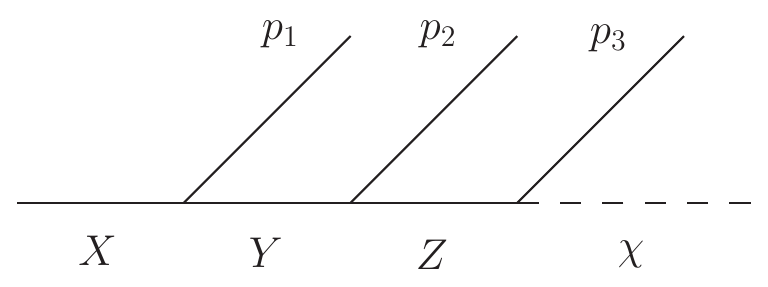}
        \caption{Feynman diagram for our benchmark decay chain. $X$, $Y$, $Z$, and $\chi$ are all new particles, while $p_{1,2,3}$ are SM particles.}
        \label{fig:FeynmanDecay}
\end{figure}

There are however final states that lie between these extremes, where algebraic methods cannot be used, but there are sufficiently many kinematic observables such that their correlations also contain crucial information that cannot be extracted by only plotting commonly used one-dimensional distributions such as kinematic edges and endpoints. It was shown in ref.~\cite{Agrawal:2013uka} that a decay chain that proceeds via three consecutive two-body decays has this property (see figure~\ref{fig:FeynmanDecay}), and subsequent papers~\cite{Debnath:2016gwz, Debnath:2018azt} explored how an analysis based on the full dimensionality of the Lorentz-invariant observables can be used to enhance discovery prospects as well as to improve the precision of mass measurements. In other words, in these event topologies there are features of the phase space distribution that only reveal themselves with a multidimensional study, but not in the standard one-dimensional projections. In particular, it was shown that the variable $\Delta_4$, to be reviewed in the next section, is a highly effective observable for measuring not only the mass differences between successive particles in the cascade decay, but also the overall mass scale along the “flat direction”, where the locations of kinematic edges and endpoints in each step of the cascade remain fixed. In this paper, we also adopt the same decay chain as the benchmark case for our study. The decay chain that we consider in this paper has been previously studied using end-points in kinematic observables~\cite{Gjelsten_2004}. The effectiveness of a number of kinematic observables for spin determination has also been studied~\cite{Cheng:2010yy, Wang_spin, Meade:2006dw, Athanasiou:2006ef}. 

\begin{figure}
    \centering
    \begin{subfigure}{0.45\textwidth}
        \includegraphics[width = \textwidth, height = 1.03\textwidth]{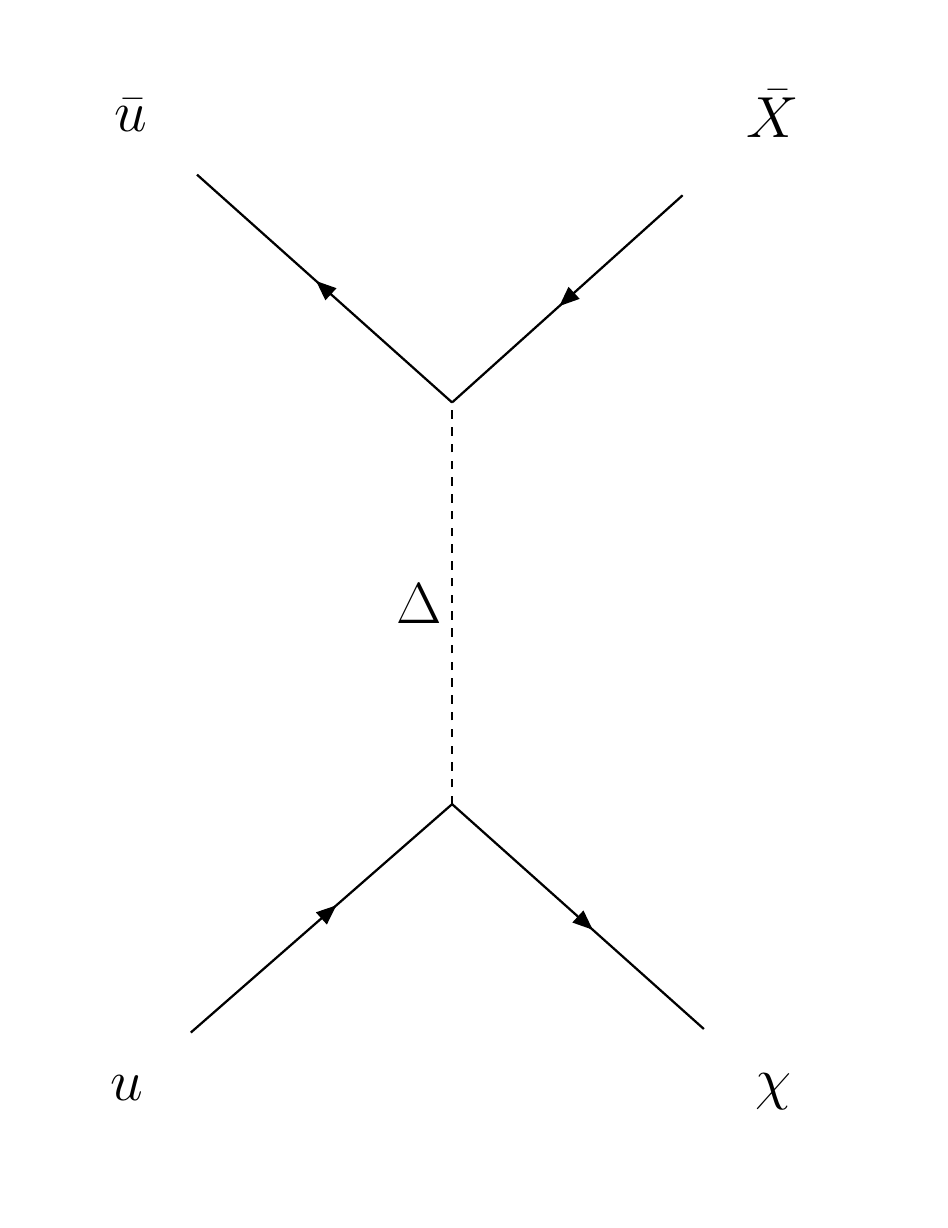}
        \caption{fermionic DM}
    \end{subfigure}
    \hfill
    \begin{subfigure}{0.45\textwidth}
        \includegraphics[width = \textwidth, height = \textwidth]{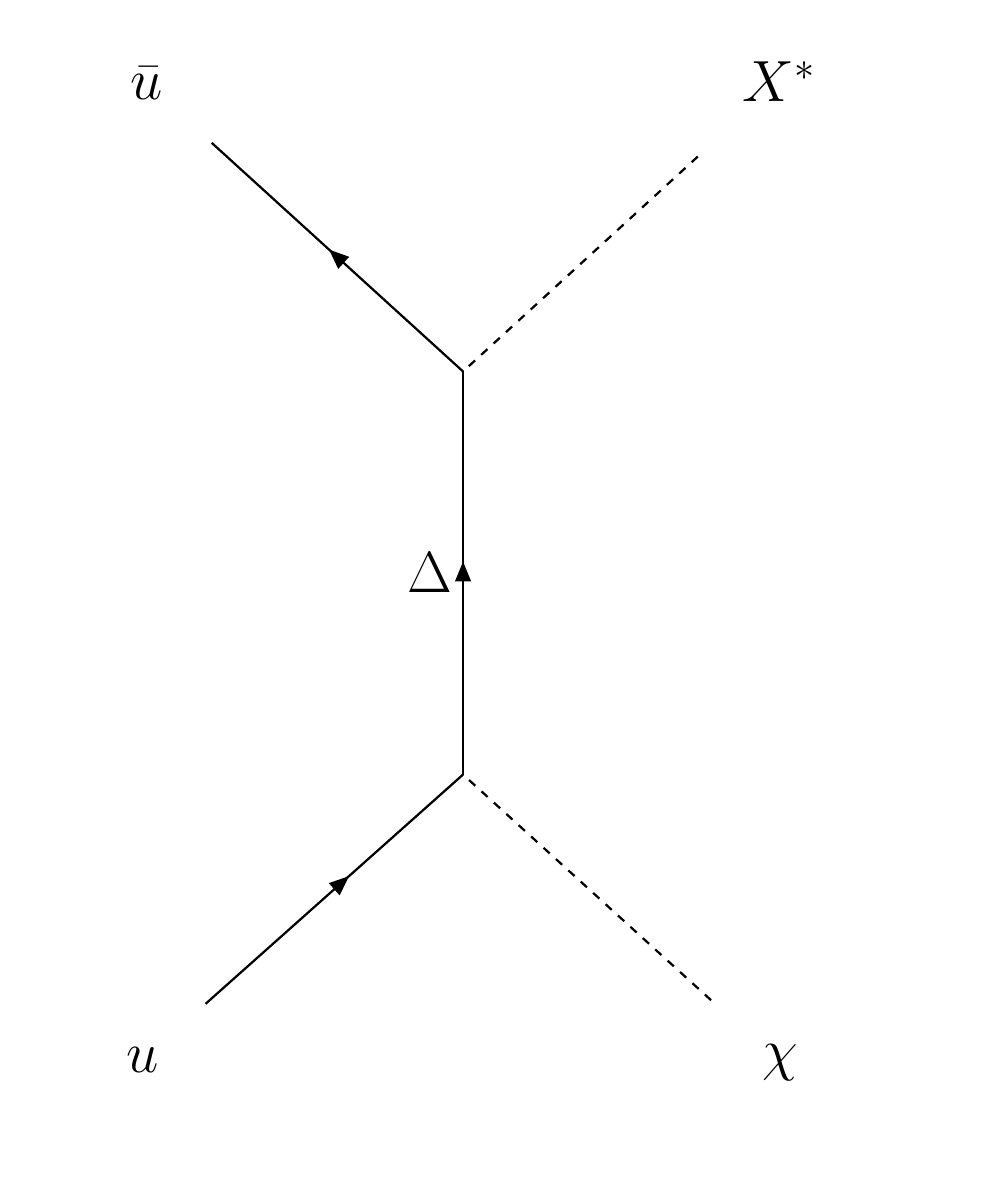}
        \caption{scalar DM}
    \end{subfigure}
    \caption{The Feynman diagrams for the production of the new particles in our benchmark event topology with the two spin configurations we consider. Fermions are represented by solid lines while scalars are represented by dotted lines.}
    \label{fig:production_feynman}
\end{figure}

Since we are considering SUSY-like collider signatures, the new particles need to be pair produced. However, if the decay chain of figure~\ref{fig:FeynmanDecay} appears on both sides of the event, then this event topology would in fact contain sufficiently many visible final state particles that algebraic methods could be used. We instead focus on scenarios where the associated production of $X$-$\chi$ is the dominant production channel (or possibly the next-to-dominant channel, after $\chi$-$\chi$ production, but for sufficiently heavy $\chi$, this channel is challenging to observe). One can for instance consider a $t$-channel production diagram, where the mediator in the $t$-channel  couples more strongly to $\chi$ than it does to $X$. This is illustrated in figure~\ref{fig:production_feynman} for two choices of particle spins, which we will introduce in section~\ref{sec:DLanalysis}. All the visible information in the event is then coming from the $X$-side of the event, and the $\chi$ on the other side of the event only affects the total MET vector among the observables.

Recently, advances in machine learning techniques have led to significant improvements in multidimensional data analysis, and in this paper we evaluate the effectiveness of these techniques for the discovery, spin determination and mass measurement in this benchmark decay chain, as a representative of SUSY-like decays that have sufficiently many kinematic observables to warrant a multidimensional analysis, but not enough for a full reconstruction via algebraic techniques. In this study, we first consider how well a deep neural network (DNN) can achieve these goals without human guidance, that is, by taking the final state momenta in the events as its only input. We then proceed to interpret the DNN in terms of human-level (HL) kinematic variables, and we show that an analysis based on the HL variables results in performance comparable to that of the DL inputs. In particular, our results confirm that $\Delta_4$ is highly correlated with the output of the DNN, in agreement with the conclusions of references~\cite{Agrawal:2013uka, Debnath:2016gwz, Debnath:2018azt}.

Supervised machine learning based analyses have been previously applied for the discovery of exotic collider events in~\cite{Baldi:2014kfa, Chang:2020rtc}. In~\cite{NN-param}, the authors have proposed methods for discovery across a range of model parameters. Applications also include resolution of combinatorics in complex topologies~\cite{Badea-combinatorics}, and parameter inference~\cite{Andreassen_2020, Brehmer:2020cvb}. 

The visible particles $p_{1,2,3}$ can be chosen among a large set of SM final states. In order to keep our analysis as simple as possible, we adopt a benchmark scenario that is as clean as possible in terms of its collider signatures. Once the proof of concept for the methods we present has been established, additional analyses can be performed to adapt these methods to more challenging final states as well. To be specific, we choose $p_{1}$ and $p_{2}$ to be same-flavor, opposite-charge leptons, such as $\mu^{\pm}$-$\mu^{\mp}$, and we choose $p_{3}$ to be a photon. This final state can arise for example if $Y$ is a muon partner, while $X$, $Z$ and $\chi$ are partners to gauge bosons. We emphasize that we are not promoting this to be a particularly plausible scenario of beyond-the-SM physics. It is only meant to provide a relatively clean example of our benchmark event topology, to study the effectiveness of the methods we are proposing. $X$, $Z$ and $\chi$ can all be fermions and $Y$ a boson, or the other way around, and we in fact study both assignments to address not only the questions of discovery and mass measurement, but of spin determination as well.

With these choices, the final state we focus on in this study is $\mu^+ \mu^- \gamma$ + MET. Since there is no reason to expect a resonant structure in the $\mu^+ \mu^-$ system in the signal, we impose a $Z$-veto in the analysis to eliminate the leading backgrounds. We also veto any significant hadronic activity. With these choices, there are two SM backgrounds contributing to this final state. The first is $Z$-$\gamma$ pair production, with the $Z$ decaying to taus, and both taus decaying to muons. The second is triboson production, such as $W^+ W^- \gamma$ (with both $W$'s decaying to muons) and $ZZ^* \gamma$ (with the on-shell $Z$ decaying invisibly and the off-shell $Z$ decaying to muons). These are the backgrounds we include in our analysis.

Given the signal and the background described above, a problem that needs to be dealt with early on is the following: Since information about the masses of the new particles and their spins is not known prior to their discovery, we cannot train the neural networks using Monte-Carlo events generated with the correct signal spectrum and the correct spin assignments. As a possible resolution to this problem, we study the effectiveness of `scanning' over the parameter space using a performance metric that quantifies discovery ($S/\sqrt{B}$ in our case). A full scan over the parameter space is computationally very expensive and we will only perform a local scan in order to demonstrate that the true spectrum is at least a local maximum.

\begin{figure}[h!]
    \centering
    \includegraphics[scale = 0.3]{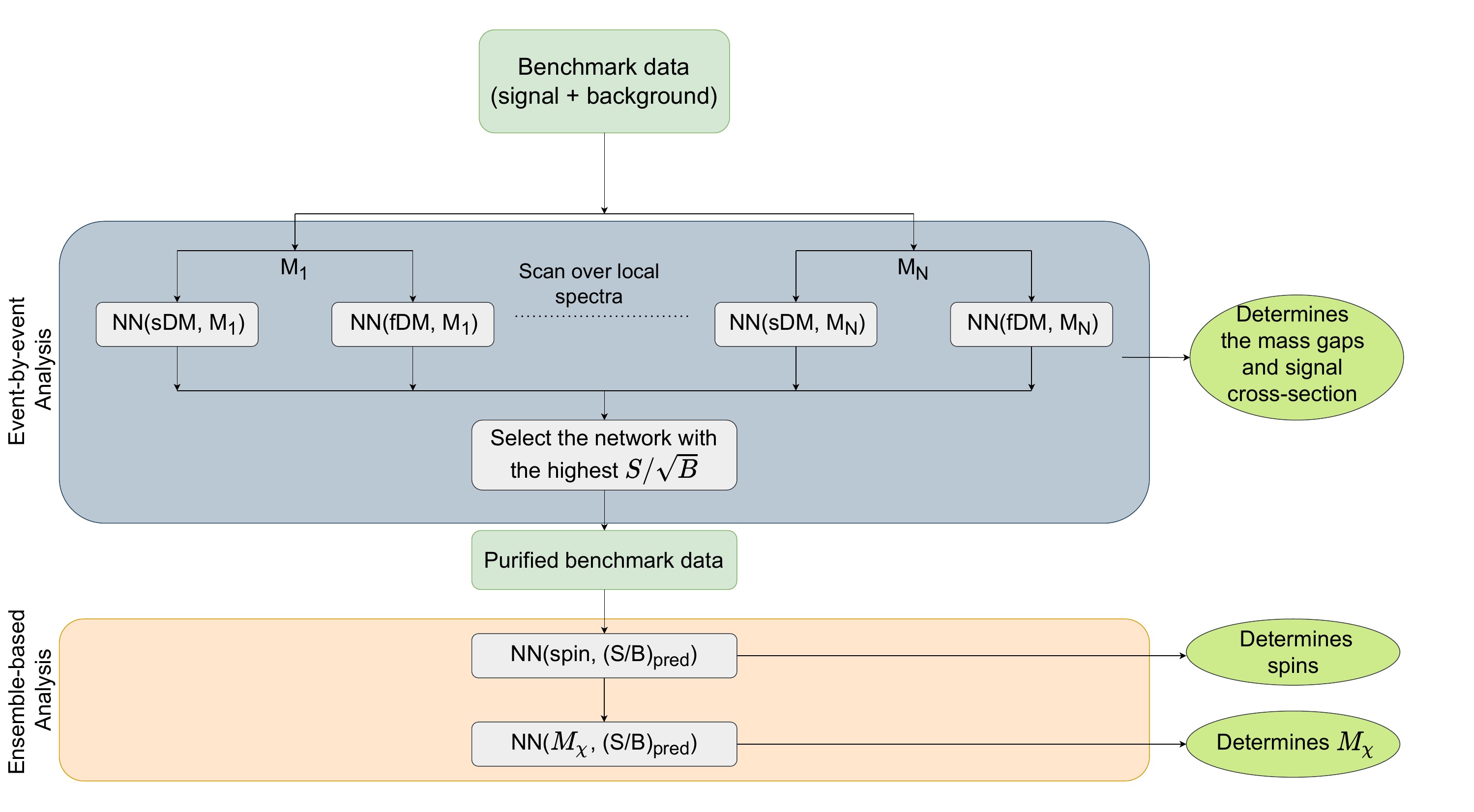}
    \caption{Flowchart detailing the overall procedure for discovery, mass gap determination, spin determination and overall mass scale determination.}
    \label{fig: flowchart}
\end{figure}

First, we perform the scan using networks trained solely on the final state momenta, which we will refer to as the Detector Level (DL) variables. We show that over the region of this local scan, the discovery metric is maximized for the networks trained using mass spectra for which the mass differences, or more precisely, the kinematic edges/ endpoints in the three $m_{ij}$ variables have the correct values. Then, we proceed to identify a set of Human Level (HL) variables that match the DNN. In particular, we show that performing the scan using the HL variables rather than the final state momenta results in an improved resolution for the mass gap measurement. As we will see, both the DL and the HL variable-based scans do not perform well in measuring the spins and the overall mass scale. We show that analyses based on ensemble methods using HL kinematic variables can be used to determine the spins and the overall mass scale. Fig~\ref{fig: flowchart} shows a schematic representation of our procedure.

This paper is organized as follows: We start with a short review of the relevant kinematic observables in section~\ref{sec:PS}, and a review of the relevant machine learning techniques in section~\ref{sec:ML}. Then in section~\ref{sec:DLanalysis}, we study the prospects for discovery and mass measurement relying on the final state momenta of the visible particles, and a deep neural network trained on individual events. We then proceed to interpret the DNN in terms of HL variables in section~\ref{sec:HLanalysis}, and we perform the second, ensemble-based, stage of the analysis in order to determine the spins of the new particles (in section~\ref{sec:spin}), and their masses (in section~\ref{sec:mass}) much more accurately. We conclude in section~\ref{sec:conclusions}.

\section{Review of Phase Space and Kinematic Observables}
\label{sec:PS}

In this section we review important properties of kinematic observables that are used in the analysis of the next few sections. Using the notation of figure~\ref{fig:FeynmanDecay}, we label the four-momenta of the new particles in a given event as $p_{X}^{\mu}$ etc, and those of the visible final states as $p_{1}^{\mu}$ etc. Very generally, the distribution of events in phase space is given by
\beq
d\Gamma=d\Pi_{PS}\, |{\mathcal M}|^2,
\eeq
where $d\Pi_{PS}$ stands for the phase space volume element, and $|{\mathcal M}|^2$ for the amplitude squared. While the latter contains valuable information about the spins of the particles, and the angular correlations of the final state particles, the information about the spectrum of the new particles and about the boundary of the kinematically available phase space is entirely contained in the phase space factor. As mentioned in the introduction, we consider two possible spin assignments for the new particles. In our analysis of discovery and mass measurement prospects, we consider both possibilities, and we look for commonalities as well as differences in the list of the optimal HL variables in these two cases.

The simplest way to characterize the data is of course through the momentum vectors $\vec{p}_{1,2,3}$ of the three visible final state particles, and the of the transverse missing energy $({\rm MET})$. As usual in collider analyses, we express these vectors in terms of $p_{\rm T}$, the pseudorapidity $\eta$ (except in the case of the MET) and the azimuthal angle $\phi$ of the final state particles. Considering boost invariance along the beam direction, differences $\Delta\eta$ and $\Delta\phi$ between any two particles are very useful observables, as is the combination $\Delta R=\sqrt{\Delta\eta^2+\Delta\phi^2}$. The fully Lorentz-invariant observables include the pair invariant masses $m_{ij}^2=\left(p_{i}^{\mu}+p_{j}^{\mu}\right)^2$, as well as the total invariant mass of all three visible particles $m_{123}^{2}$. One less well known Lorentz-invariant observable is $\Delta_{4}$ which we define below. This completes the list of kinematic observables that we consider in our paper.

The maximum values of the $m_{ij}^{2}$ and $m_{123}^{2}$ variables for this final state topology are well known~\cite{Allanach:2000kt, Gjelsten_2004, Gjelsten:2005aw, Miller:2005zp, Matchev:2009iw, Burns:2009zi}, and often used for mass (difference) measurements in this event topology. They are given by:
\begin{align}
  (m_{23}^2)_{\rm max}
  &=
  (M_Y^2-M_Z^2)(M_Z^2-M_{\chi}^2)/M_Z^2,
  \label{eq:endpoint23}
\end{align}
\begin{align}
  (m_{12}^2)_{\rm max}
  &=
  (M_X^2-M_Y^2)(M_Y^2-M_Z^2)/M_Y^2,
  \label{eq:endpoint12}
\end{align}
\begin{align}
  (m_{13}^2)_{\rm max}
  &=
  (M_X^2-M_Y^2)(M_Z^2-M_{\chi}^2)/M_Z^2 ,
  \label{eq:endpoint13}
\end{align}
\begin{align}
  (m_{123}^2)_{\rm max}
  &=
  \begin{cases}
    \frac{(M_X^2-M_Y^2)(M_Y^2-M_{\chi}^2)}{M_Y^2 }
    & \frac{M_X}{M_Y} > \frac{M_Y}{M_Z} \frac{M_Z}{M_{\chi}}, \\
    \frac{(M_X^2 M_Z^2 - M_Y^2 M_{\chi}^2)
    (M_Y^2-M_Z^2)}
    {M_Y^2 M_Z^2}
    & \frac{M_Y}{M_Z} > \frac{M_Z}{M_{\chi}} \frac{M_X}{M_Y}, \\
    \frac{(M_X^2-M_Z^2)(M_Z^2-M_{\chi}^2)}{M_Z^2 }
    & \frac{M_Z}{M_{\chi}} > \frac{M_X}{M_Y} \frac{M_Y}{M_Z}, \\
    (M_X - M_{\chi})^2
    & \mathrm{otherwise}.
  \end{cases}
  \label{eq:endpoint123}
\end{align}
It is straightforward to see that these formulas are sensitive to differences of masses, however there exists a flat direction along which the masses can be varied such that the endpoints of all $m^2_{ij}$ variables remain constant. The kinematic variable $\Delta_4$ we are about to describe  has been shown to be useful for measurements of the mass spectrum along this challenging direction~\cite{Agrawal:2013uka, Debnath:2016gwz, Debnath:2018azt}.

For each decaying $X$ particle, the momenta of the final states, $p^{1,2,3}$ and $\chi$ can be represented as a point in 4-body phase space. While it is not commonly used, there is an elegant description of 4-body phase space~\cite{Byers:1964ryc} that is manifestly Lorentz-invariant. Consider the $4\times4$ matrix ${\mathcal Z}_{ij}$ whose elements are given by $p_{i}\cdot p_{j}$ (where we are taking $\chi$ to be the fourth particle). Define the functions $\Delta_i$ of these momenta as the coefficients in the characteristic polynomial of ${\mathcal Z}$, namely
\beq
{\rm Det}\left[\lambda I_{4\times 4} - {\mathcal Z}\right]\equiv \lambda^4 - \lambda^3 \Delta_1 - \lambda^2 \Delta_2 - \lambda \Delta_3 - \Delta_4.
\eeq
The kinematically allowed region in phase space for $X$ decay can be shown~\cite{Byers:1964ryc} to be defined by the conditions $\Delta_{i} > 0$ for all $i=1,\dots,4$, with the boundary of the region defined by $\Delta_4 = 0$ (with $\Delta_{1,2,3}$ still positive). As already mentioned, $\Delta_4$ has been shown to be a powerful observable for analyzing this decay chain, both for discovery and for mass measurement purposes~\cite{Agrawal:2013uka, Debnath:2016mwb, Debnath:2016gwz, Debnath:2018azt,  Altunkaynak:2016bqe}. There is one subtlety that needs to be mentioned. Using the definition above, calculating $\Delta_4$ requires knowledge of all final state momenta, including that of $\chi$, which is invisible. However, if the masses $\left\{ M_X , M_Y , M_Z, M_{\chi}\right\}$ in the spectrum are assumed known, then all dot products $p_{i} \cdot p_{\chi}$ (and therefore $\Delta_4$) can be calculated from only the visible particle momenta, by using the on-shell conditions for the intermediate particles.

The reason that $\Delta_4$ is such a useful variable is that the volume element of 4-body phase space, expressed in the differentials $dm_{ij}^2$, is given by $\Delta_4^{-1/2}$, up to a constant and an energy-momentum conserving delta function. As a result, in the Lorentz-invariant coordinates $m^2_{ij}$, the distribution of signal events is strongly clustered near $\Delta_4=0$. Background events on the other hand do not arise from $X$ decays, and their $\Delta_4$ distribution has no similar sharp feature at $\Delta_4=0$. As a result, if the spectrum were somehow known, then an excess of signal events over background can be easily discovered due to the sharp peak in the $\Delta_4$ distribution near zero. Of course, discovery must precede a knowledge of the spectrum, however $\Delta_4$ can be used to accomplish both goals simultaneously. In principle, one can scan over the possible spectra $\left\{ M_X , M_Y , M_Z, M_{\chi}\right\}$, and look for an excess near $\Delta_4=0$. This feature will be most significant when the correct spectrum is used, so the presence of the excess will both serve as a discovery variable, and as a way to measure the unknown particles masses. Of course, in practice it is computationally prohibitive to perform a scan in all four mass variables. Fortunately, the well-known kinematic edges and endpoints in the $m^2_{ij}$ distributions already provide good sensitivity for the mass differences, and therefore they can be used first, leaving only the overall mass scale undetermined (parametrized by $M_{\chi}$, say) along the flat direction. Then, one can perform a one-dimensional scan over $M_{\chi}$ and use $\Delta_4$ to fix the spectrum completely. Our analysis in the rest of this paper will demonstrate that neural network techniques based on $\Delta_4$ do indeed result in a high precision measurement of the masses along the flat direction.

\section{Review of Machine Learning Tools}
\label{sec:ML}

 For the results presented here, all neural networks were implemented using the Keras~\cite{Chollet2015keras} package with the TensorFlow~\cite{tensorflow2015-whitepaper} backend. Wherever we implement neural networks in our analysis, we provide details about the number of nodes, layers and activation functions within the respective section.
 
 In the initial phase of our analysis focused on discovery, we implement fully connected deep networks as binary classifiers, the two target classes being the signal and the background. We use the binary cross-entropy as the loss function. 

Unlike the first stage of the analysis described in sections~\ref{sec:DLanalysis} and~\ref{sec:HLanalysis}, where the neural network analyzes one event at a time and assigns each event an output in the range [0,1] where 0 corresponds to background and 1 corresponds to signal, we will see in sections~\ref{sec:spin} and~\ref{sec:mass} that the spin determination and the measurement of the overall mass scale are more challenging, and an  ensemble-based analysis is required. In the ensemble-based analysis, the entire data-set is considered as a single input. We also identify the HL variables whose distributions are particularly effective in separating the different hypotheses. Histograms in these variables are then used for the training and evaluation of the networks. Specific details about the histogram binning are provided in the respective sections of the paper. 

In our case, post-discovery, the determination of the mass spectrum reduces to determining a single mass value along the flat direction, which we parameterize by $M_\chi$. We treat this as a regression problem and we use the Mean Squared Error loss function for training the network. 

The Average Decision Ordering (ADO) metric introduced in \cite{Whiteson} is very useful for quantifying the correlation between two functions, one of which may be the neural network output and the other an analytic function defined in terms of the inputs. ADO is constructed to quantify the degree to which two functions $f$ and $g$ rank pairs of events belonging to the two classes A and B in the same order. We use the following discrete version of the ADO:
\begin{equation}
\text{ADO}[f,g]=\sum_{x \in A}\sum_{x' \in B}\Theta \Big( \big( f(x) - f(x') \big) \big( g(x) - g(x') \big) \Big),
\end{equation}
where $\Theta$ is the Heaviside function.

\section{Analysis Based Solely on Final State Momenta}
\label{sec:DLanalysis}

In this section, we start by studying how well a `black box' DNN can discriminate between signal and background based on detector level variables alone, by which we mean the momenta of the final state particles, in other words without using any guidance in the form of human level variables. Since the DL variables represent all the available information at the detector, the DNN learns an approximation of the optimal discriminating function, namely the ratio of signal and background distributions. In the next section, we will identify a small combination of HL variables that are sufficient to approximate the optimal discriminator, relying on ADO as a metric. Finally, in sections~\ref{sec:spin} and~\ref{sec:mass}, we will combine the strength of these HL variables with ensemble-based analysis to tackle the more challenging problems of spin determination and the measurement of the overall mass scale. Figure~\ref{fig: flowchart} provides a visual overview of the stages of the analysis.

We treat the signal cross section essentially as a free parameter varying in a range consistent with the new particles having electroweak couplings and masses of a few hundred GeV. In order to work on a specific example, we choose the following signal spectrum as a benchmark: 
\begin{equation}
M_X = 390~{\rm GeV},\quad M_Y = 360~{\rm GeV},\quad M_Z = 330~{\rm GeV}, \quad M_\chi = 300~{\rm GeV}.     
\label{eq:truth_spectrum}
\end{equation}
This will be denoted as the {\it truth spectrum} for the rest of the paper. For the background cross sections, we use the leading order values obtained from Monte Carlo simulations.

For the spin assignments of the new particles, we work with two possibilities, denoting these as the fDM and sDM models, based on whether $\chi$ is a fermion or boson (see figure~\ref{fig:production_feynman}). In these two models, the new particles are taken to be:
    
\begin{itemize}

\item fDM: $X, Z, \chi$ are neutral fermions while $Y$ is a charged scalar.

\item sDM: $X, Z, \chi$ are neutral scalars while $Y$ is a charged fermion.

\end{itemize} 
We begin by describing the details of Monte Carlo methods we use in our analysis.

\subsection{Monte Carlo methods and selection cuts}
\label{sec:MC_and_cuts}
    
As mentioned in the introduction, the production mechanism of interest to us is $pp\rightarrow X \chi$. We take the production to proceed via a heavy t-channel mediator $\Delta$ (which is a scalar/fermion for the fDM/sDM signal model). We take this mediator to couple to up-type quarks. This choice is mostly arbitrary, and motivated by the fact that the constraints on new physics from flavor violation are weaker compared to new particles coupling to down-type quarks. Since the analysis below is based entirely on the decay of $X$, and we treat the signal cross section as a free parameter, with our results parameterized by this parameter. For the background, we generate $pp \rightarrow \mu^+ ~\mu^- ~\gamma ~\nu_l ~\bar{\nu}_l$ and $pp \rightarrow \mu^+ ~\mu^- ~\gamma ~\nu_l ~\bar{\nu}_l~\nu_l ~\bar{\nu}_l$ events, with all possible neutrino flavor combinations. These contain the dibson and triboson processes (including off-shell $W$/$Z$-bosons) discussed in the introduction. Signal and background events are all generated using MadGraph~\cite{Boos:2001cv}. 
   
The final state can be described by 9 momentum components of the 3 observable particles. We remind the reader that any significant hadronic activity will be vetoed - as a result, the MET does not carry additional information to the momenta of the visible particles. There is a combinatorial ambiguity in the final state since $p_1$ can be a $\mu^+$ and $p_2$ a $\mu^-$ or the other way around (see figure~\ref{fig:FeynmanDecay}). Therefore it is useful to denote the final state momenta by their charges, namely $p_+$, $p_-$, and $p_\gamma$. 
    
In Monte Carlo generation, we use relatively loose cuts, demanding only $p_{T,\gamma} > 4 \text{GeV}$ to avoid singularities in the matrix element, and we impose the acceptance cut of $|\eta|<2.5$. In order to simulate the detector energy resolution for muons and photons, we use the  parameters estimated by ATLAS for the high luminosity run. For muons, these are given by~\cite{CERN-LHCC-2017-017}
    
    \begin{equation}
        {\frac{\sigma(E)}{E}} = {1.61\times 10^{-2}} \oplus {2.76\times 10^{-3}\text{GeV}^{-1/2}{\sqrt{E}}},
    \end{equation}
    
    and for photons they are given by~\cite{CERN-LHCC-2017-018}
    
    \begin{equation}
        {\frac{\sigma(E)}{E}} = {9.84 \times 10^{-3}} \oplus {\frac{9.41 \times 10^{-2}\text{GeV}^{1/2}}{\sqrt{E}}} \oplus \frac{1.19\text{GeV}}{E}.
    \end{equation}

There are two (non-prescaled) ATLAS triggers that are relevant for our final state: a dimuon trigger requiring $p_T > 14$~GeV for both muons, and a $\mu\mu\gamma$ trigger requiring $p_T > 10$~GeV for both muons and $p_T > 15$~GeV for the photon. We include both trigger paths in our event selection. Since our main interest in this paper is in signal spectra with small mass splittings, the invariant mass values take on relatively small values, and a Z-veto can be applied to the $\mu^+$-$\mu^-$ system to eliminate the large backgrounds with on-shell $Z$'s. In order to also avoid the increased background at low invariant mass due to photon conversions, we impose the selection cut
\begin{equation}
    15 \text{GeV} < m_{+-} < 65 \text{GeV}.
\end{equation}
After all selection cuts, the background cross section is given by 2.88~fb.

We will work with event samples that correspond to the full HL-LHC luminosity of $3000~{\rm fb}^{-1} $, and we report on the performance of the analysis as a function of the signal strength, quantified by S/B, where S and B are defined as the number of signal and background events after the selection cuts listed above. 


\subsection{Measurement of mass differences}
\label{sec:DLdeltam}

We start our analysis using a DNN trained on DL variables. As mentioned before, the questions of discovery and mass measurement are linked since the DNN needs to be trained without prior knowledge of the spectrum. We will see below that the first phase of our analysis is efficient in measuring the mass differences between the new particles, but not the overall mass scale. 

As mentioned in section~\ref{sec:intro}, it is beyond our computational resources to scan over all possible spectra. We settle for a less ambitious goal of at least demonstrating that the correct mass differences present an optimal point in a {\it local} scan of a `testing spectrum' ($\vect{M_0}$) given by
\begin{equation}
\vect{M_0} = (M_X, M_Y, M_Z, M_\chi) = (691, 660, 631, 600)~{\rm GeV},
\label{eq:testingspec}
\end{equation}
which has the correct mass differences, but is displaced from the truth spectrum (equation~\ref{eq:truth_spectrum}) along the flat direction, which corresponds approximately to the vector $(1,1,1,1)$ in the $(M_X, M_Y, M_Z, M_\chi)$ space. We set up the local scan around this testing spectrum, using the following orthogonal basis:
\begin{align*}
     \vect{v_1} = \frac{1}{2}(0,1,0,-1), \hspace{2.5ex} \vect{v_2} = \frac{1}{{2}}(1,0,-1,0), \hspace{2.5ex} \vect{v_3} = \frac{1}{2}(1,-1,1,-1), \hspace{2.5ex} \vect{v_4} \approx (1,1,1,1),
\end{align*}
where $\vect{v_4}$ is taken along the true flat-direction, which deviates slightly from the vector (1,1,1,1). The spectra over which we perform the scans are parametrized as:

\begin{equation}
    \vect{M} = \vect{M}_0 + (\alpha \vect{v_1} + \beta \vect{v_2} + \gamma \vect{v_3} + \delta \vect{v_4})
    \label{eq:localscan}
\end{equation}

 For each of the spectra $\vect{M}$, we generate two sets of Monte-Carlo signal events: one with the spins assigned according to the fDM hypothesis and the other according to the sDM hypothesis. Then, we train an ensemble of DNNs, two for each $\vect{M}$, corresponding to the two possible spin assignments, to distinguish the respective signal from the background. Based on the events used to train the network, we refer to these networks as the fDM network and the sDM network respectively.

Each DNN is made up of 3 hidden layers containing 200 hidden nodes. Nodes in the intermediate layers are assigned ReLu activation functions, and the output node is assigned the sigmoid activation function. The input layer contains 9 nodes corresponding to the $3 \times 3 = 9$ observable momentum components. Additionally, we implement an early stopping monitor, which has a patience of 20 epochs on the validation set. Each network is trained using 1M events each for the signal and the background.
 
Once the DNNs are trained, we study their output on the benchmark sample (signal at the truth spectrum plus background). In order to infer the dependence of the performance on the signal cross-section, we list our results below for three values of S/B, namely 1.0, 0.1 and 0.01.

\begin{figure}
    \centering
    \includegraphics[width = 0.65\textwidth]{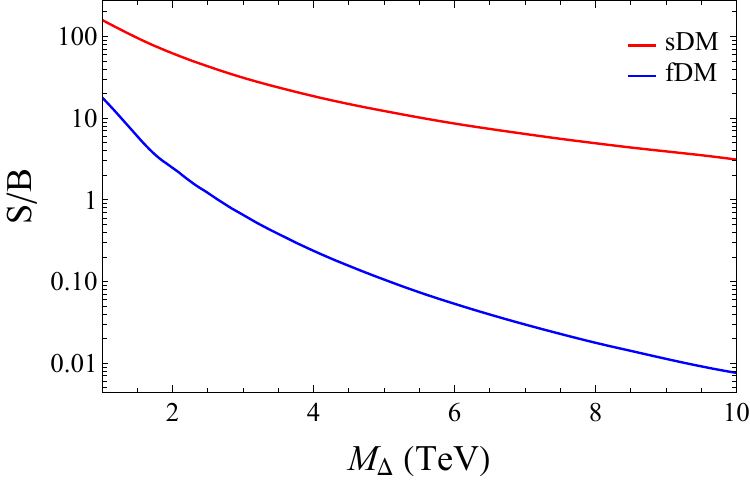}
    \caption{Cross-sections for the sDM(red) and fDM(blue) signal (at the truth spectrum) as a function of the mediator mass. All relevant couplings are taken to be 1.}
    \label{fig:cs}
\end{figure}

In order to provide context for these values of S/B, figure~\ref{fig:cs} shows how S/B depends on the mass of the mediator ($M_\Delta$) in the sDM and fDM spin assignments, with the couplings at all vertices of figure~\ref{fig:production_feynman} taken to have the numerical value 1.

\begin{figure}[h!]
    \centering
    \begin{subfigure}{0.45\textwidth}
    \includegraphics[width = \textwidth]{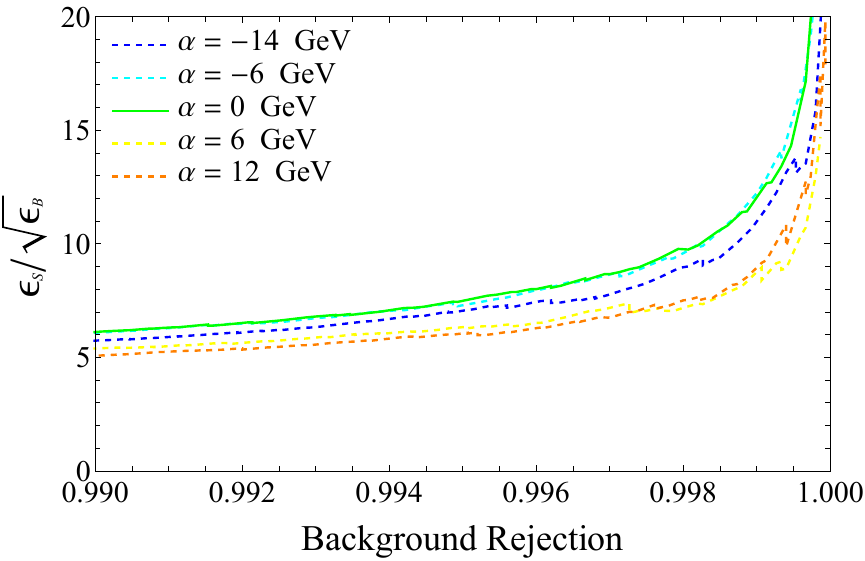}
    \caption{$\alpha$}
    \end{subfigure}
    \hfill
    \begin{subfigure}{0.45\textwidth}
    \includegraphics[width = \textwidth]{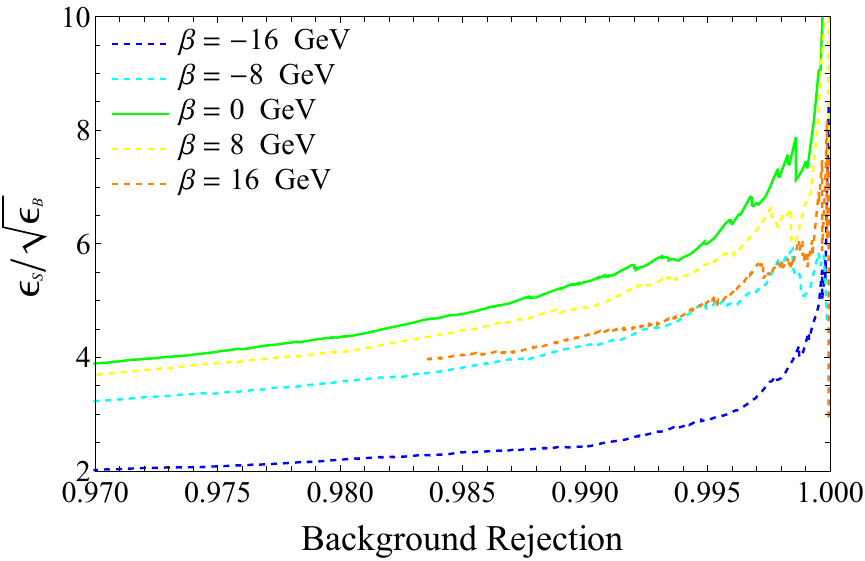}
    \caption{$\beta$}
    \end{subfigure}
    \hfill
    \begin{subfigure}{0.45\textwidth}
    \includegraphics[width = \textwidth]{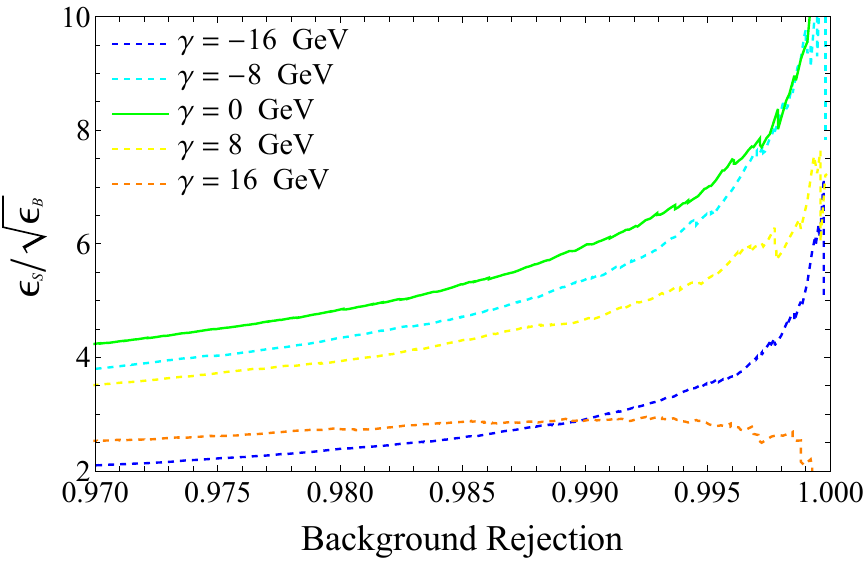}
    \caption{$\gamma$}
    \end{subfigure}
    \hfill
    \begin{subfigure}{0.45\textwidth}
    \includegraphics[width = \textwidth]{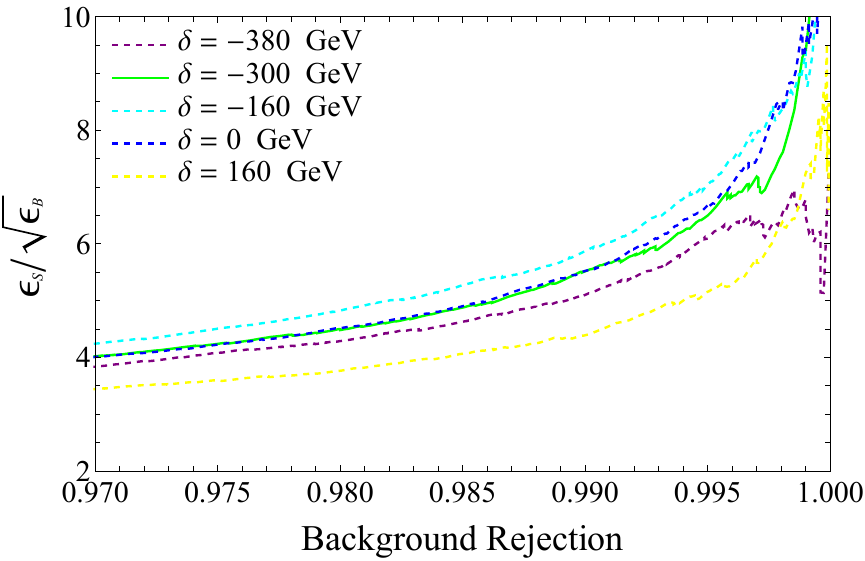}
    \caption{$\delta$}
    \end{subfigure}
    \caption{For the local mass scan around the testing spectrum with S/B = 1.0, the parameters $\alpha$, $\beta$ and $\gamma$ favor the correct mass differences in the spectrum, and stronger cuts give rise to better performance. In contrast, the scan along the flat direction shows no preference for the true value of the overall mass scale. The discovery cut we impose corresponds to a background rejection of 0.995. In each panel, the solid green contour corresponds to the true value of the parameter.}
    \label{fig: example-scan}
\end{figure}

We quantify the performance by $\epsilon_S/\sqrt{\epsilon_B}$, defined as the enhancement in $S/\sqrt{B}$ after applying a cut on the output of the network, compared to the selection cuts of section~\ref{sec:MC_and_cuts}. The choice of $\epsilon_S/\sqrt{\epsilon_B}$ (as opposed to $\epsilon_S / \epsilon_B$) as the performance metric is motivated by the expectation that statistical uncertainties will dominate over systematic ones. In figure~\ref{fig: example-scan}, we plot $\epsilon_S/\sqrt{\epsilon_B}$ for S/B = 1.0. It can be seen that the performance continues to improve towards stronger cuts on the network output. Here we only plot the metric with the correct spin assignment for simplicity. However, the trend remains the same even if the incorrect spin assignment is used. In order to preserve sufficient signal statistics for the later stages of the analysis, in each network, we apply a cut on the network output that results in $\epsilon_B = 0.005$ (this tends to correspond to $\epsilon_S \approx 0.5$ for the best performing networks in our analysis). Hereafter, this will be referred to as the ``discovery cut''. The value of the discovery cut is inferred from Monte Carlo samples at each mass point. With the discovery cut, the minimum value of $S/B$ such that the signal statistical significance is boosted above $5\sigma$ with the discovery cut is $(S/B)_{\rm min}\sim 6.7\times 10^{-3}$. We will therefore focus on the range $S/B > 0.01$ for the rest of our analysis.

We perform one-dimensional scans using the parametrization of equation~\ref{eq:localscan}. The winning mass hypothesis is chosen to be the one that results in the highest $\epsilon_S/\sqrt{\epsilon_B}$ on the benchmark sample, with the discovery cut, setting $\epsilon_B$ to 0.005. To account for statistical uncertainties, we run 50 iterations of the scan. From the example scan in figure~\ref{fig: example-scan}, it can be seen that this procedure results in an accurate measurement of the  parameters $\alpha$, $\beta$ and $\gamma$, thereby fixing the mass-gaps. In contrast, we find that the $\epsilon_S/\sqrt{\epsilon_B}$ metric (at a fixed background rejection) fluctuates randomly as $\delta$ is varied, and therefore the value of $\delta$ maximizing the metric is uncorrelated with the true value.

To most efficiently utilize our computing resources, we limit the local mass scan to the following range around the testing spectrum:
$\alpha, \beta, \gamma \in (-20, 20) ~\text{GeV}$ and $\delta \in (-500, 500) ~\text{GeV}$, with spacings of $\Delta\alpha = \Delta\beta = \Delta\gamma = 0.5 ~\text{GeV}$ and $\Delta\delta = 20 ~\text{GeV}$. Note that with these choices, the mass hierarchy $M_X > M_Y > M_Z > M_\chi$ is automatically preserved.

\begin{table}[h!]
    \centering
    \begin{subtable}{\textwidth}
    \centering
    \begin{tabular}{c c c c }
    \hline
    S/B & $\alpha$ & $\beta$ & $\gamma$\\
    \hline
    1.0 & $-1.6 \pm 1.4$ & $5.5 \pm 0.4$ & $1.5 \pm 0.3$ \\
    
    0.1 & $-3.7 \pm 3.3$ & $4.7 \pm 2.2$ & $-0.1 \pm 1.9$\\
    
    0.01 & $-4.0 \pm 9.2$ & $3.4 \pm 5.7$ & $-0.4 \pm 4.5$\\
    \hline
    \end{tabular}
    \caption{sDM}
    \end{subtable}
    
    \bigskip
    
    \begin{subtable}{\textwidth}
    \centering
    \begin{tabular}{c c c c }
    \hline
    S/B & $\alpha$ & $\beta$ & $\gamma$\\
    \hline
    1.0 & $-12.4 \pm 3.4$  & $4.8 \pm 1.1$ & $-4.8 \pm 2.2$ \\
    
    0.1 & $-10.1 \pm 6.0$ & $4.1 \pm 2.3$ & $-3.2 \pm 3.0$ \\
    
    0.01 & $-9.0 \pm 9.9$ & $4.0 \pm 8.0$ & $-2.4 \pm 5.5$\\
    \hline
    \end{tabular}
    \caption{fDM}
    \end{subtable}
    
    \caption{The average values and uncertainties of the scan parameters obtained over 50 scan iterations along each direction (all numbers in GeV).}
    \label{tab:scans-DL}
\end{table}

Table~\ref{tab:scans-DL} summarizes the results of the scans (averaged over the scan iterations). As expected, we find that the mass differences are determined to a precision of a few GeV by this procedure, while the overall mass scale is left essentially unconstrained. Note that for high values of $S/B$ there is a bias in the results for $\alpha$, $\beta$ and $\gamma$. As we will see in the next section, when HL variables are used, these biases are eliminated.

\section{Identification of Optimized Variables}
\label{sec:HLanalysis}

After the DNN analysis based on the DL inputs, our next goal is to search for HL variables which match the performance of this analysis. We employ the ADO-guided method prescribed in ref.~\cite{Whiteson} for identifying the set of HL variables that have the highest correlation with respect to the DL network. The definition of ADO was given in section~\ref{sec:ML}. For our purposes here, the ADO is calculated from the fraction of pairs of events (one drawn from the signal generated at the testing spectrum and the other drawn from the background) which are ranked in the same order by both the deep network operating on the DL inputs and a given set of HL variables. 

To summarize the method presented in ref.~\cite{Whiteson}, one picks the HL variable with the highest ADO with respect to the DL network (say $f_1$) in the first iteration. In the next iteration, only the pairs of events for which $f_1$ and the DL network result in dissimilar orderings are considered. The second HL variable $f_2$ is then picked to be the one with the highest ADO over these subset of pairs and so on. We terminate the process when the additional HL variable does not lead to a significant improvement in the AUC.

As we described in section~\ref{sec:PS}, we will consider the following list of HL variables:

\begin{itemize}
    \item The transverse momenta of the three visible final state particles : $(p_{T+},p_{T-},p_{T\gamma})$
    \item The transverse missing energy : MET
    \item The invariant masses of the three visible final state particle pairs : $(m_{+-}, m_{+\gamma}, m_{-\gamma})$
    \item The total invariant mass of the three visible final state particles : $m_{+-\gamma}$
    \item $\Delta_4$ computed assuming the testing spectrum.
    \item $\Delta R_{ij} = \sqrt{\Delta\eta_{ij}^2+\Delta\phi_{ij}^2}$ for all pairs of final state particles.
\end{itemize}

\begin{table}[h!]
\centering
\begin{subtable}{0.5\linewidth}
\centering
 \begin{tabular}{c  c} 
 \hline
 variable & ADO \\ [0.5ex] 
 \hline 
 MET & 0.835 \\
 
 $\Delta_4$ & 0.812 \\
 
 $p_{T_\gamma}$ & 0.724  \\
 
 $p_{T_+}$ & 0.702 \\ 
 
 $p_{T_-}$ & 0.699 \\
 
 $m_{-\gamma}$ & 0.631\\
 
 $m_{+\gamma}$ & 0.631\\
 
 $m_{+-\gamma}$ & 0.622\\
 
 $m_{+-}$ & 0.568\\
 \hline
 \end{tabular}
\caption{sDM network}
\label{table: sdm-map}
\end{subtable}%
\hfill
\begin{subtable}{0.5\linewidth}
\centering
 \begin{tabular}{c  c} 
 \hline
 variable & ADO \\ [0.5ex] 
 \hline 
 MET & 0.880 \\
 
 $\Delta_4$ & 0.827 \\
  
 $p_{T_+}$ & 0.739 \\
 
 $p_{T_-}$ & 0.721 \\
 
 $p_{T_\gamma}$ & 0.711  \\
 
 $m_{-\gamma}$ & 0.617\\
 
 $m_{+\gamma}$ & 0.632\\
 
 $m_{+-}$ & 0.558\\
 
 $m_{+-\gamma}$ & 0.552\\
 \hline
 \end{tabular}
\caption{fDM network}
\label{table: fdm-map}
\end{subtable}
\caption{HL variables with the leading ADOs for the sDM and the fDM networks. The ADOs have been computed over pairs of signal events at the  testing spectrum and the background. We consider 20,000 events each of signal and background.}
\end{table}

\begin{figure}[h!]
     \centering
     \begin{subfigure}[b]{0.45\textwidth}
         \centering
         \includegraphics[width=\textwidth]{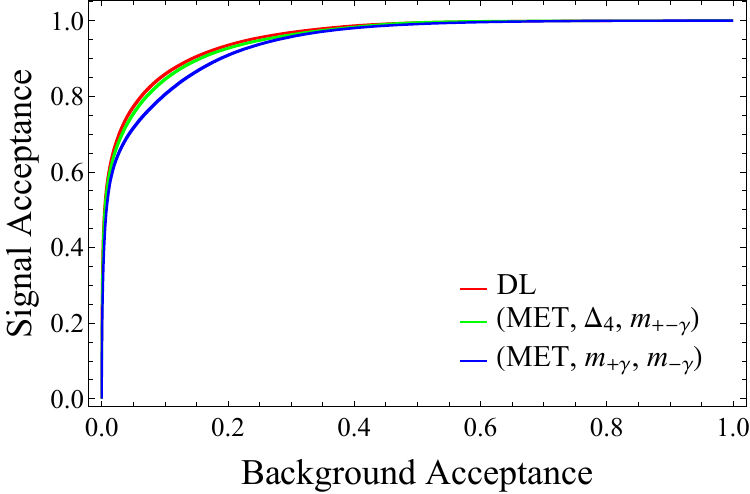}
         \caption{sDM}
         \label{fig: ROC scalar}
     \end{subfigure}
     \hfill
     \begin{subfigure}[b]{0.45\textwidth}
         \centering
         \includegraphics[width=\textwidth]{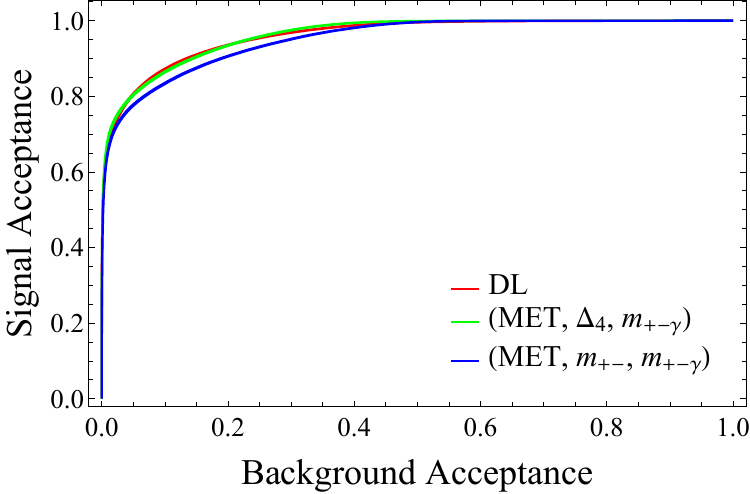}
         \caption{fDM}
         \label{fig:ROC fermion}
     \end{subfigure}
        \caption{ROC curves for both spin models, for the DNN trained DL inputs (in red) and the HL variables (in green). For comparison, we also present the performance of the leading triplet of HL variables that does not contain $\Delta_4$ (in blue).}
        \label{fig: ROC}
\end{figure}

\begin{table}[h!]
    \centering
    \begin{subtable}{0.5\linewidth}
    \centering
    \begin{tabular}{c c}
    \hline
    Variables  &  AUC\\ [0.5ex]
    \hline
    DL & 0.9686\\
    
    \hline
    (MET, $\Delta_4$, $m_{+-\gamma}$)     & $0.9618$ \\
    
    (MET, $\Delta_4$, $m_{+-}$)     & $0.9572$ \\
    
    (MET, $\Delta_4$, $m_{+\gamma}$)     & $0.9401$ \\
    \hline
    (MET, $m_{+-}$, $m_{+-\gamma}$) & 0.9448 \\
    
    (MET, $m_{+\gamma}$, $m_{-\gamma}$) & 0.9401 \\
    
    (MET, $m_{+-\gamma}$, $m_{+\gamma}$) & 0.9361 \\
    \hline
    \end{tabular}
    \caption{sDM}
    \end{subtable}%
    \hfill
    \begin{subtable}{0.5\linewidth}
    \centering
    \begin{tabular}{c c}
    \hline
    Variables  &  AUC\\ [0.5ex]
    \hline
    DL & 0.9580\\
    \hline
    (MET, $\Delta_4$, $p_{T_\gamma}$)     & $0.9505$ \\
    
    (MET, $\Delta_4$, $m_{+-\gamma}$)     & $0.9492$ \\
    
    (MET, $\Delta_4$, $m_{+-}$)     & $0.9445$ \\
    \hline
    (MET, $m_{+\gamma}$, $m_{-\gamma}$) & 0.9285 \\
    
    (MET, $m_{+-}$, $m_{+-\gamma}$) & 0.9265\\
    
    (MET, $m_{+\gamma}$, $m_{+-\gamma}$) & 0.9187\\
    \hline
    \end{tabular}
    \caption{fDM}
    \end{subtable}%
    \caption{Triplets of HL variables with and without $\Delta_4$ having the highest AUCs over the benchmark data-set. For simplicity, we only list the numbers for when the training data set uses the correct spin model.}
    \label{tab: AUC}
\end{table}

\begin{figure}[h!]
     \begin{subfigure}[b]{0.3\textwidth}
         \centering
         \includegraphics[width=\textwidth]{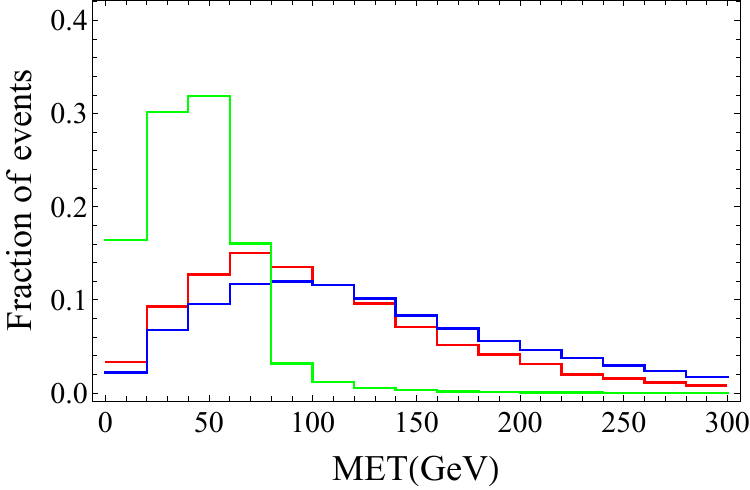}
         \caption{MET}
         \label{fig: met}
     \end{subfigure}
     \hfill
     \begin{subfigure}[b]{0.3\textwidth}
         \centering
         \includegraphics[width=\textwidth]{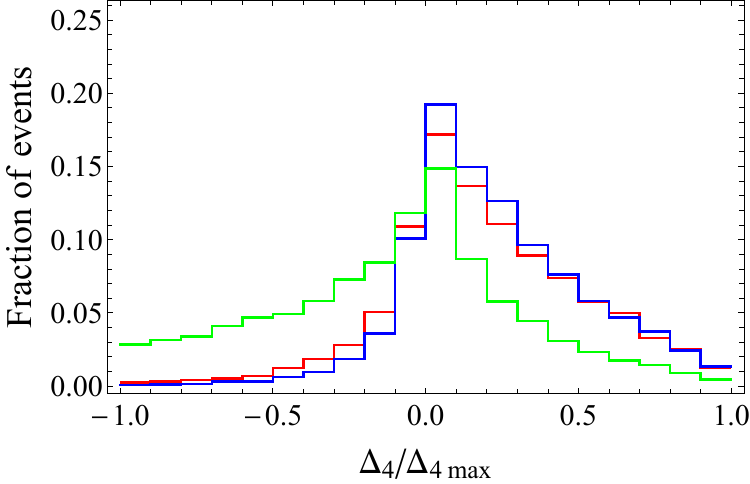}
         \caption{$\Delta_4$}
         \label{fig: delta4}
     \end{subfigure}
     \hfill
     \begin{subfigure}[b]{0.3\textwidth}
         \centering
         \includegraphics[width=\textwidth]{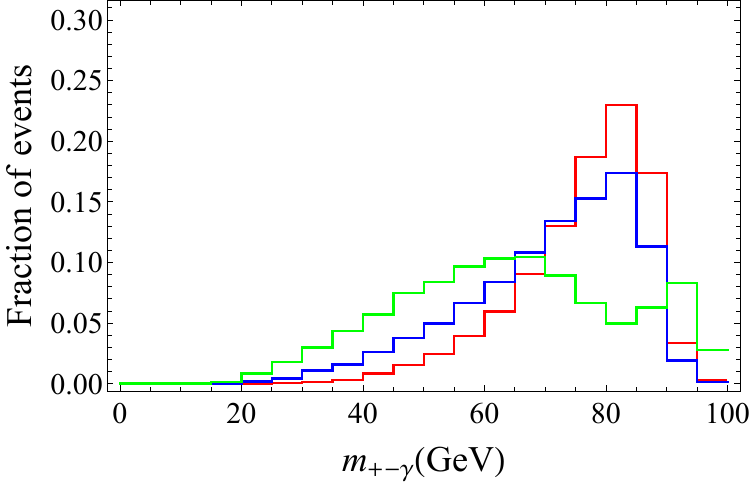}
         \caption{$m_{+-\gamma}$}
         \label{fig: m123}
     \end{subfigure}
      
        \caption{Distributions of the leading triplet of HL variables for pure benchmark signal in the sDM (red) and fDM (blue) models, and for background (green). $\Delta_4$ is calculated using the testing spectrum.}
        \label{fig: hist}
\end{figure}

Tables~\ref{table: sdm-map} and~\ref{table: fdm-map} show the ADOs of individual HL variables for the two spin models generated at the testing spectrum. The AUC that is obtained by combining triplets of these variables is listed in table~\ref{tab: AUC}. The HL networks have 3 hidden layers containing 50 nodes each. The combination of the variables $(\text{MET}, \Delta_4, m_{+-\gamma})$ is the most effective when both spin models are considered, and it effectively matches the AUC of the DNN based on DL inputs. We also illustrate this visually in figure~\ref{fig: ROC}, showing the performance of the HL variables (MET, $\Delta_4$, $m_{+-\gamma}$) (in green) and the DL-based neural network (in red). For comparison, we also show the leading HL triplet that does not include $\Delta_4$ (in blue). Our results confirm the power of $\Delta_4$ as a kinematic variable in analyzing this decay chain, in accordance with prior studies. In figure~\ref{fig: hist}, we show the distributions of the HL variables for the two signal models and for the background.

\begin{figure}[h!]
    \centering
    \begin{subfigure}{0.45\textwidth}
    \includegraphics[width = \textwidth]{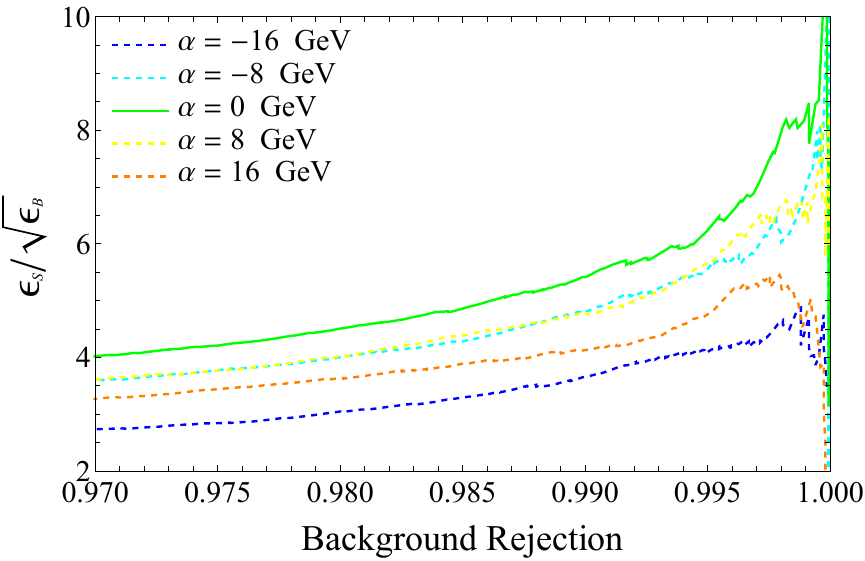}
    \caption{$\alpha$}
    \end{subfigure}
    \hfill
    \begin{subfigure}{0.45\textwidth}
    \includegraphics[width = \textwidth]{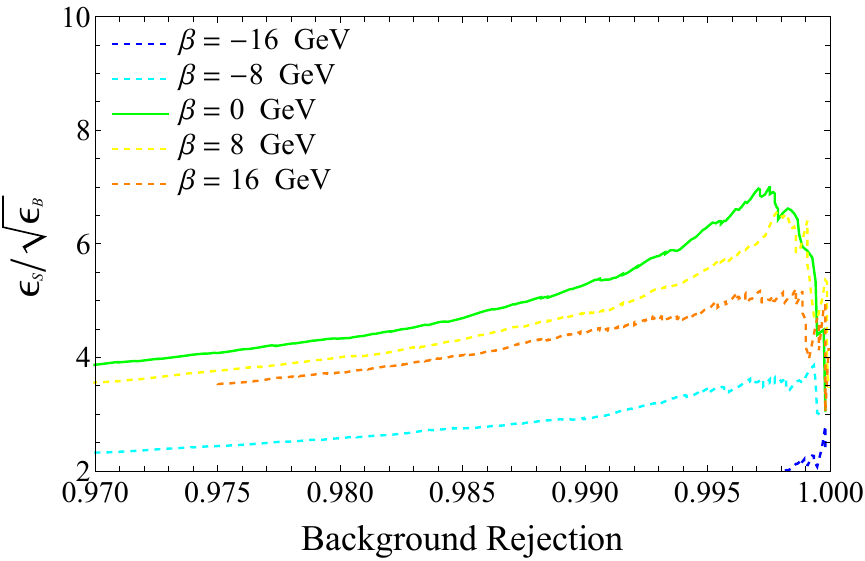}
    \caption{$\beta$}
    \end{subfigure}
    \hfill
    \begin{subfigure}{0.45\textwidth}
    \includegraphics[width = \textwidth]{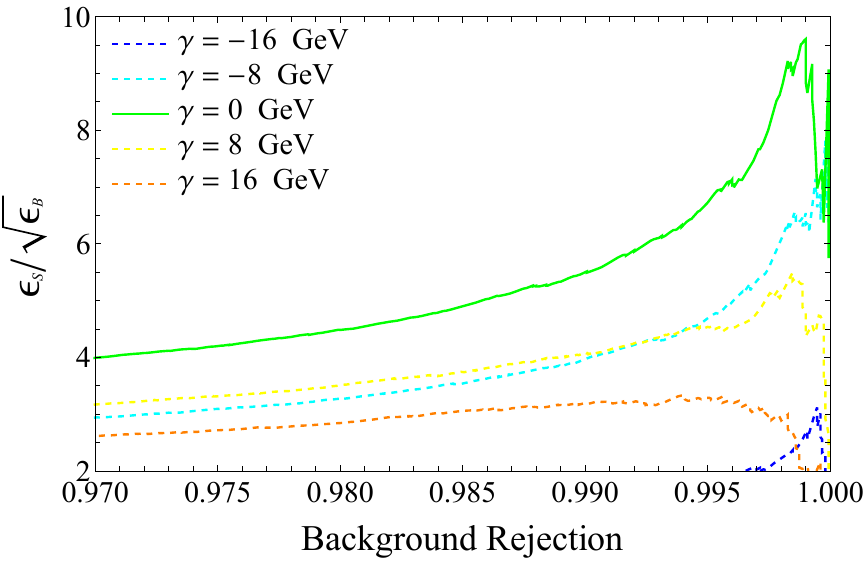}
    \caption{$\gamma$}
    \end{subfigure}
    \hfill
    \begin{subfigure}{0.45\textwidth}
    \includegraphics[width = \textwidth]{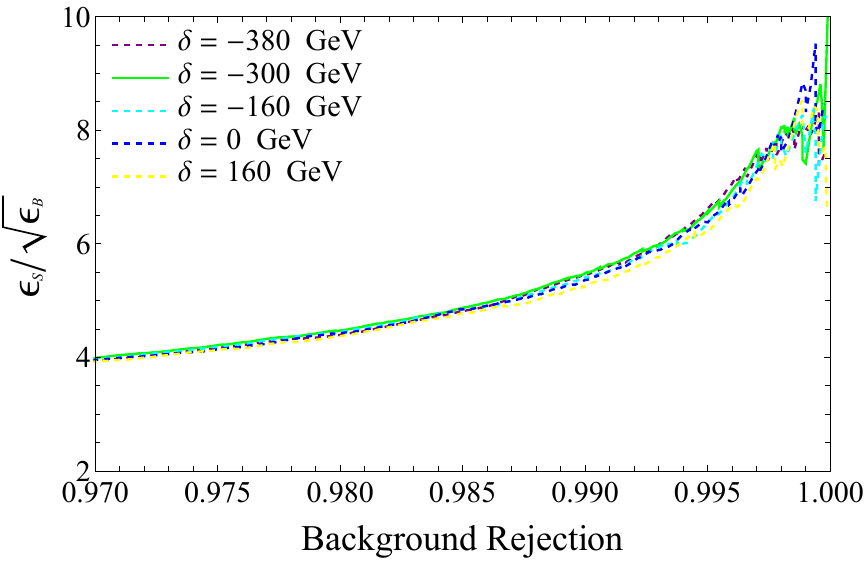}
    \caption{$\delta$}
    \end{subfigure}
    \caption{Local scans around the testing spectrum as in figure~\ref{fig: example-scan}, but using the leading triplet of HL variables.}
    \label{fig: example-scan-HL}
\end{figure}

Next, we turn our attention to studying the performance of the HL variables for determining the spectrum, namely fixing the parameters $\alpha, \beta, \gamma, \delta$. Figure~\ref{fig: example-scan-HL} is the counterpart of figure~\ref{fig: example-scan}, made using the leading triplet of HL variables. Similarly, table~\ref{tab:scans-HL} is the counterpart of table~\ref{tab:scans-DL}. 

\begin{table}[h!]
    \centering
    \begin{subtable}{\textwidth}
    \centering
    \begin{tabular}{c c c c c }
    \hline
    S/B & $\alpha$ & $\beta$ & $\gamma$\\
    \hline  
    1.0 & $-0.7 \pm 0.3$ & $1.2 \pm 0.5$ & $-0.2 \pm 0.2$\\
    0.1 & $-0.6 \pm 1.3$ & $1.1 \pm 0.9$  & $-0.5 \pm 0.8$\\
    0.01 & $2.3 \pm 6.0$  & $2.0 \pm 2.7$  & $-0.5 \pm 2.9$\\
    \hline
    \end{tabular}
    \caption{sDM}
    \end{subtable}
    
    \bigskip
    
    \begin{subtable}{\textwidth}
    \centering
    \begin{tabular}{c c c c c }
    \hline
    S/B & $\alpha$ & $\beta$ & $\gamma$\\
    \hline
    1.0 & $0.0 \pm 0.8$ & $1.3 \pm 0.3$ & $-0.6 \pm 0.8$\\
    0.1 & $0.0 \pm 1.5$ & $0.9 \pm 0.8$ & $-0.8 \pm 0.8$ \\
    0.01 & $1.6 \pm 7.1$ & $3.0 \pm 3.2$ & $-1.3 \pm 5.4$ \\
    \hline
    \end{tabular}
    \caption{fDM}
    \end{subtable}
    
    \caption{The average values and uncertainties of the scan parameters (all numbers in GeV), as in table~\ref{tab:scans-DL}, using the leading triplet of HL variables.}
    \label{tab:scans-HL}
\end{table}

Given the relatively large dimensionality of the DL inputs, combined with finite training samples and network parameters, it is hard for the DL network to converge towards the global minimum of the loss function. Within the limitations of our analysis, we find that identification of the optimal set of HL variables leads to a better convergence towards the global minimum. This is evident from comparing the resolution for mass differences (Tables~\ref{tab:scans-DL} and~\ref{tab:scans-HL}). The overall mass scale however still remains unresolved. In the next two sections, we will further improve the performance of the neural networks in order to determine the spins of the new particles and measure the overall mass scale at a much higher precision. In order to achieve this goal, we will first purify the signal in the data by passing the data through the HL classifier network (trained at the testing spectrum), and apply the discovery cut, which eliminates $99.5$\% of background events in each network. Since this corresponds to a roughly $50\%$ efficiency for signal events, it results in an enhancement in S/B by a factor of $\sim 100$.

\begin{figure}[h!]
     \centering
     \begin{subfigure}[b]{0.45\textwidth}
         \centering
         \includegraphics[width=\textwidth]{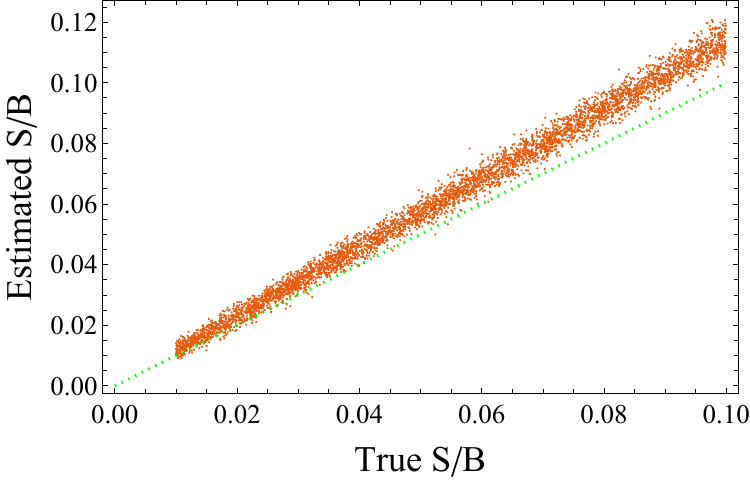}
         \caption{sDM}
         \label{fig: CS scalar}
     \end{subfigure}
     \hfill
     \begin{subfigure}[b]{0.45\textwidth}
         \centering
         \includegraphics[width=\textwidth]{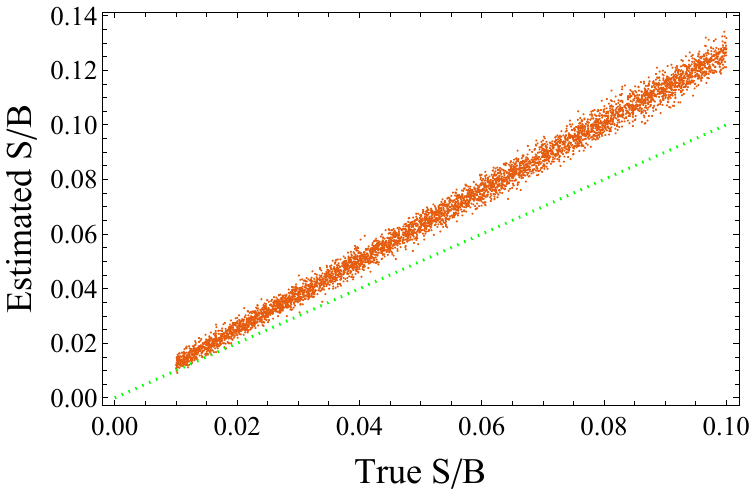}
         \caption{fDM}
         \label{fig: CS fermion}
     \end{subfigure}
        \caption{The estimated value of $S/B$ for the benchmark sDM and fDM data samples (red points), plotted against the true $S/B$ (green curve), based on the formula of equation~\ref{eq: cs-det}. The results were obtained over 5000 pseudo-experiments in each case, using the HL networks.}
        \label{fig: CS-det}
\end{figure}

We can also estimate S/B, without prior knowledge of the overall mass scale. This will be used in the second stage of the analysis, where we employ ensemble-based methods. We have

\begin{equation}
    (S/B)_{\rm est.} = \frac{\epsilon_{\rm S+B}- \epsilon_{\rm B}}{\epsilon_{\rm S}-\epsilon_{\rm S+B}},
    \label{eq: cs-det}
\end{equation}

where $\epsilon_{S}$ is evaluated in the network with the highest $\epsilon_{S}/\sqrt{\epsilon_{B}}$ after applying the discovery cut, using signal events generated at the testing spectrum. $\epsilon_{\rm S+B}$ is the efficiency of the discovery cut on the actual data. In other words, $\epsilon_B$ and $\epsilon_S$ are parameters obtained from simulations, while $\epsilon_{S+B}$ is measured from data. Figure~\ref{fig: CS-det} shows the accuracy of this procedure. 

\section{Spin Determination}
\label{sec:spin}

We now turn our attention to determining the spins of the new particles. We treat this as a binary classification problem. We start by introducing an ensemble-based analysis method which will also allow us to determine the overall mass scale in the next section. With the ensemble-based method, we assess the performance of DL variables first, and we then identify the HL variables that result in a similar performance.

\subsection{Ensemble-based method}
 Consider binary classification of ensembles containing $N$ events each, $\{x_i\} \in \mathbb{E}^N$, where each event $x_i \in \mathbb{E} = \mathbb{R}^k$. In our case, the ensemble corresponds to the entire data-set. A neural network performing this classification task must have $k\times N$ nodes in the input layer. Also, one must take into account the permutational invariance of the input vectors (in other words, the ordering of events within an ensemble should not affect the outcome). This can be achieved either by having a special architecture for the neural network or by considering all the allowed permutations of the input vectors during training. Given the size of the event samples in our case and the dimensionality of the final state ($k = 9$ for DL inputs), we will not attempt to do this. We instead apply the method introduced in ref.~\cite{Thaler-ensemble}, to build an ensemble-based classifier based on an event-by-event classifier. We first build a simple event-by-event classifier operating on events $x_i$ represented by the complete set of DL variables. Then, using the output of the event-by-event classifier $y(x_i)$, the ensemble-based classification function $y_N(\{x_i\})$ is generated:
\begin{equation}
    y_N(\{x_i\}) = \frac{\prod_{i}y(x_i)}{\prod_{i}y(x_i) + \prod_{i}(1 - y(x_i))}.
    \label{eq: spin}
\end{equation}

\subsection{Performance with DL variables}

The event-by-event network is made up of 3 hidden layers containing 100 nodes each. We use 1M events of each spin model to train the networks. Since our analysis so far has not allowed us to measure the overall mass scale, we continue generating the training samples at the testing spectrum. The network is trained using the labels $\hat{y}(x_i) = 1$ if $x_i \in \text{sDM}$ and $\hat{y}(x_i) = 0$ if $x_i \in \text{fDM}$. We use 100k ensembles of each spin model.

\begin{figure}[h!]
         \centering
         \includegraphics[width=0.65\textwidth]{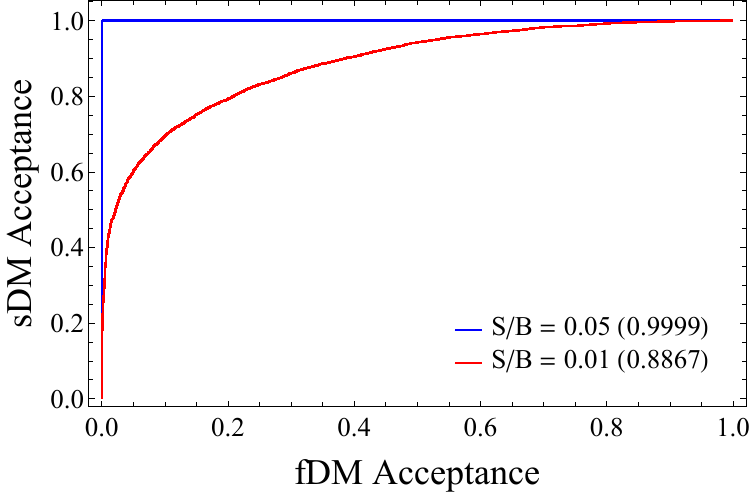}
        \caption{ROC curves for the ensemble-based spin determination using DL variables. The given S/B numbers denote the ratio before the first stage of the analysis. The corresponding AUCs are given in brackets.}
        \label{fig: spin-DL}
\end{figure}

We remind the reader that we are performing the spin determination analysis after the first stage of the analysis described in the previous section has already been performed. With the amount of signal purification we gained by applying the discovery cut, even in the most conservative cases ($S/B \gsim 0.01$ needed for discovery), the signal fraction in the events passing the cut has been boosted to an ${\mathcal O}(1)$ number. In Figure~\ref{fig: spin-DL}, we present the ROC curves for representative values of S/B, for an integrated luminosity of 3~ab$^{-1}$. As can be seen in that figure, the spin determination is very accurate and only starts to degrade near the lowest $S/B$ values of interest.

\subsection{Performance with HL variables}

Next, we identify the HL variables that match the performance of the event-by-event DL spin determination network, and then use those to perform the ensemble-based analysis in terms of these variables.   

\begin{table}[h!]

    \begin{subtable}{0.5\linewidth}
    \centering
    \begin{tabular}{c c}
    \hline
    variable & ADO\\
    \hline
     $m_{+-}$    &  0.721\\
     $\Delta R_{+-}$ & 0.663 \\
     MET & 0.656 \\
      $m_{+-\gamma}$   & 0.631\\
      \hline
    \end{tabular}
    \caption{}
    \end{subtable}%
    \begin{subtable}{0.5\linewidth}
    \centering
    \begin{tabular}{c c }
    \hline
    variables & AUC \\
    \hline
        DL & 0.6727\\
        $m_{+-}$ & 0.6487\\
        ($m_{+-}$, MET) & 0.6614\\
        \hline
    \end{tabular}
    \caption{}
    \end{subtable}
    \caption{(a) HL variables with the highest ADOs for spin determination. (b) Comparison of AUCs (for an event-by-event network) for the DL and HL inputs for spin determination. The values shown are computed for pure signal sDM and fDM events passing the discovery cut.}
    \label{tab:ado-spin}
\end{table}

\begin{figure}[h!]
         \centering
         \includegraphics[width=0.65\textwidth]{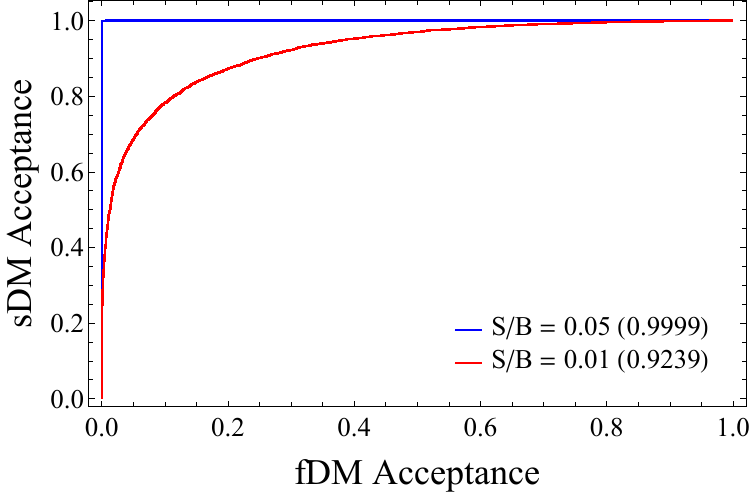}
        \caption{Same as figure~\ref{fig: spin-DL}, but using the HL variables ($m_{+-}$, MET).}
        \label{fig: spin-HL}
\end{figure}

As before, we compute the ADOs of the HL variables (at the testing spectrum). The variables with the highest ADOs for spin determination are shown in table~\ref{tab:ado-spin}. Replacing the DL variables with $(m_{+-}, \text{MET})$ results in a similar performance for event-by-event spin determination. $m_{+-}$ is identified by the ADO method to be very effective in spin determination, matching the results presented in ref.~\cite{Wang_spin}. 

\begin{figure}[h!]
     \centering
     \begin{subfigure}[b]{0.45\textwidth}
         \centering
         \includegraphics[width=\textwidth]{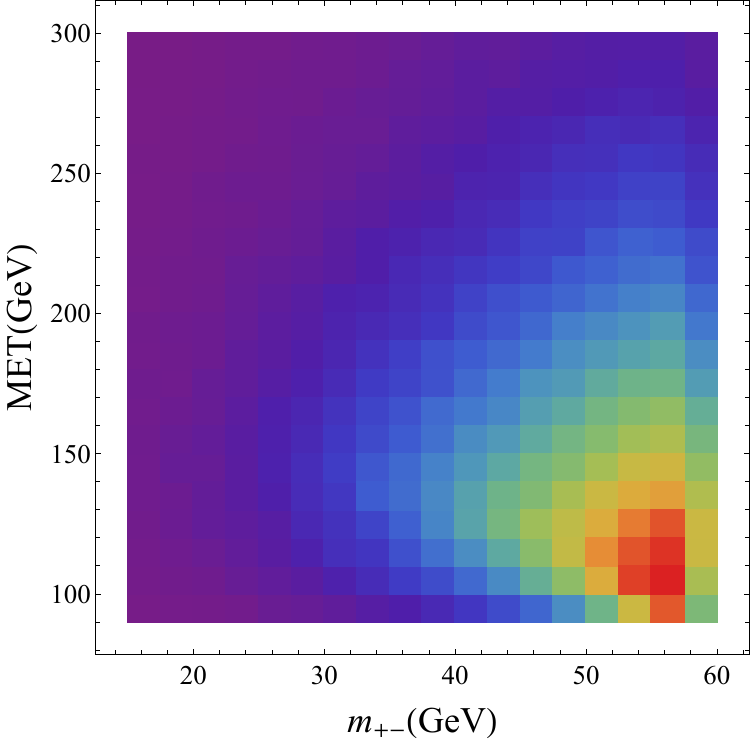}
         \caption{sDM}
         \label{fig: spin scalar}
     \end{subfigure}
     \hfill
     \begin{subfigure}[b]{0.45\textwidth}
         \centering
         \includegraphics[width=\textwidth]{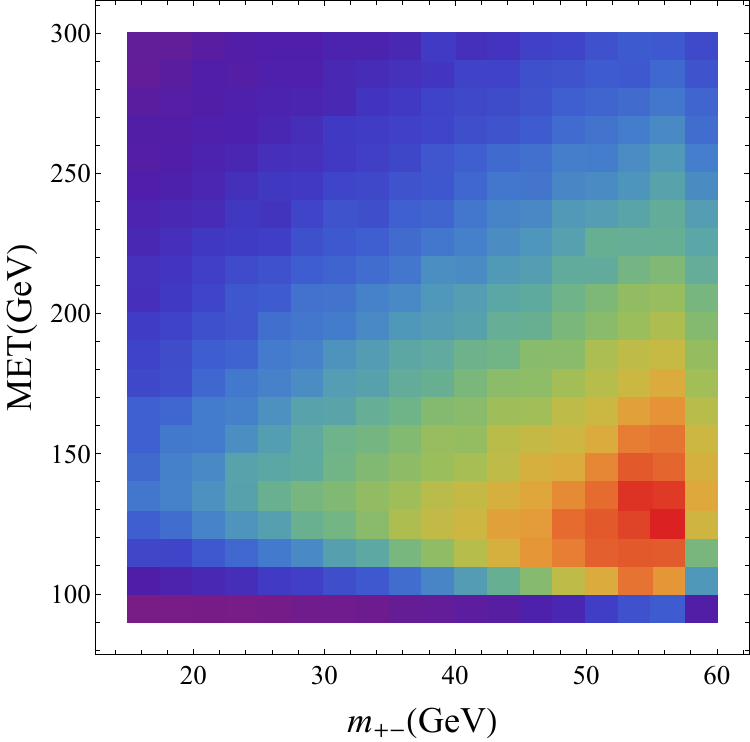}
         \caption{fDM}
         \label{fig: spin fermion}
     \end{subfigure}
        \caption{Distributions of the sDM and fDM signal events after the discovery cut in the ($m_{+-}$,MET) plane.}
        \label{fig: spin-dist}
\end{figure}

The HL network we use has two hidden layers containing 30 units each. Using the output of this event-by-event HL classifier network, we construct ensemble-based classifiers as we did for the DL variables. Figure~\ref{fig: spin-HL} shows the performance of the HL ensemble-based classifiers. Comparing figures~\ref{fig: spin-DL} and~\ref{fig: spin-HL} shows that with an ensemble-based analysis, just the two HL variables $(m_{+-}, \text{MET})$ are sufficient to match the performance of the DL network. 

\begin{figure}[h!]
         \centering
         \includegraphics[width=0.65\textwidth]{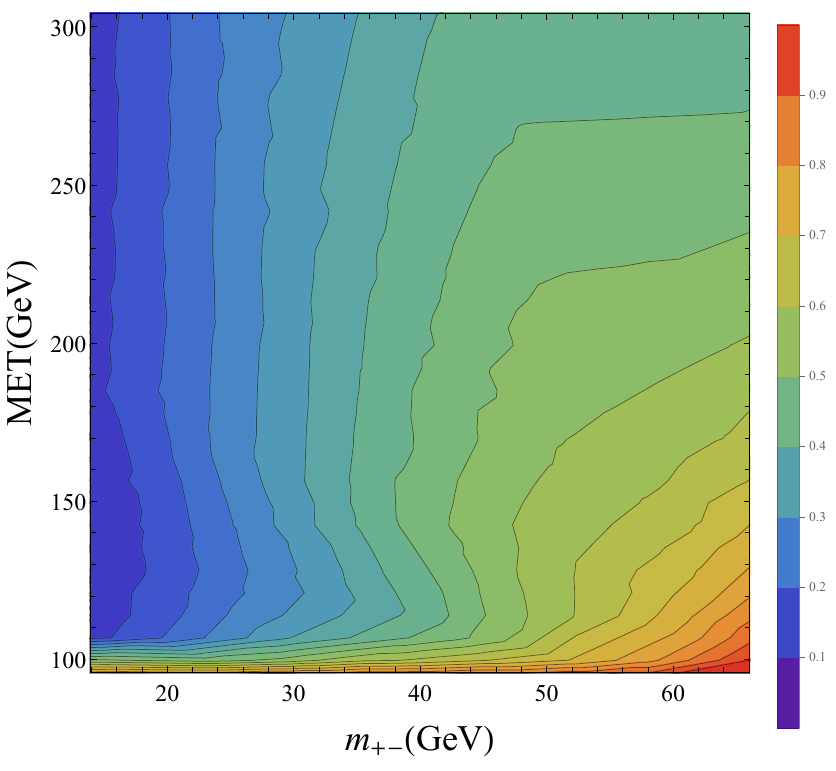}
        \caption{Contours of the output of spin determination network based on the HL variables ($m_{+-}$, MET). An output of 0 corresponds to fDM-like events, and an output of 1 corresponds to sDM-like events.}
        \label{fig: spin-contour}
\end{figure}

It is worth spending some time looking into how the spin information is encoded in the variables $m_{+-}$ and MET. In figure~\ref{fig: spin-dist}, we show the distribution of events in these two variables within the sDM and fDM signal models after the discovery cut. Note that the discovery cut eliminates almost all events with MET$<100\,$GeV, as that region of phase space is background-dominated. We also show in figure~\ref{fig: spin-contour} the contours of the event-by-event HL network in the ($m_{+-}$, MET) space. The shape of these contours can be understood with the following observations. The matrix element squared of the decay process $X\rightarrow \mu^{+}\mu^{-}\gamma\chi$ is proportional to a factor of $m_{+-}^2$ in the sDM model, but not in the fDM model, resulting in a significant difference in the $m_{+-}$ distributions in the two models, which can also be seen in figure~\ref{fig: spin-dist}. In fact, at low $m_{+-}$, the network output is basically purely based on $m_{+-}$, as can be seen on the left side of figure~\ref{fig: spin-contour}. As $m_{+-}$ approaches its maximum value, most of the kinetic energy from the decay in the $X$-frame is taken by the muons, leaving the $\chi$ with little kinetic energy. As a result, there is a boost imbalance between this softer $\chi$ produced in the $X$-decay, and the one produced directly from the initial state and recoiling against the $X$ (see figure~\ref{fig:production_feynman}). In this region, the MET and $m_{+-}$ variables are correlated, which leads to the contours on the right side of figure~\ref{fig: spin-contour} bending towards the diagonal.

\subsection{Optimizing the cut thresholds}

For the ensemble-based method described above, AUC quantifies the separation between the two spin models achieved by the output function $y_N(\{x_i\})$. However, we are ultimately interested in finding the optimal value of the cut on the output that maximizes the classification accuracy. Note that the HL variables of interest do not include $\Delta_4$, it is therefore sufficient to use the correct mass gaps and the overall mass scale is not needed. This being the case, we can use the testing spectrum for the determination of the optimal cut values.

\begin{figure}[h!]
         \centering
         \includegraphics[width=0.65\textwidth]{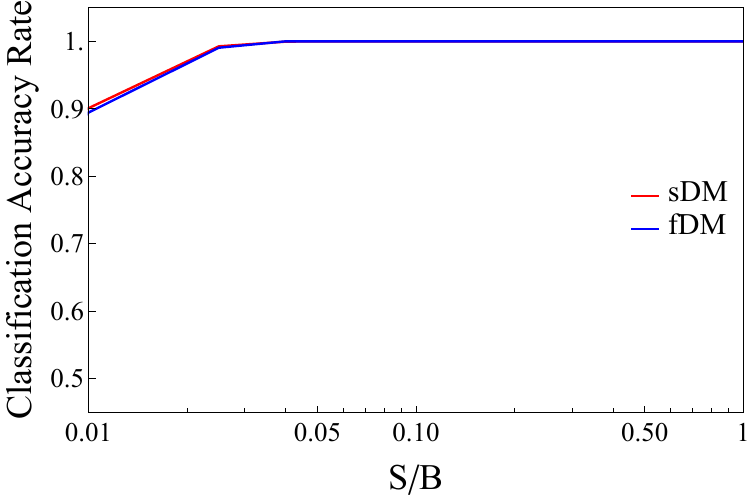}
        \caption{Accuracy rates for the spin determination using the ensemble classifier built from the event-by-event HL network.}
        \label{fig: spin-acc}
\end{figure}

For a given cut $y'_N$ on the output function $y_N(\{x_i\})$, we define the overall accuracy rate (AR) as the sum of the accuracy rates of sDM and fDM ensembles, as a function of $S/B$. Then, scanning through the range of the output function, the cut $y^*_N$ that maximizes the overall AR is found. The accuracy rates for spin determination of the benchmark signal, using HL variables, are shown in Figure~\ref{fig: spin-acc}.

\section{Mass Scale Determination}
\label{sec:mass}

Having fixed the mass differences in the earlier stage of the analysis (section~\ref{sec:HLanalysis}), the spectrum hypotheses between which we are trying to distinguish at this stage are labeled in terms of a single variable, which we can take to be $M_\chi$. It was shown in ref.~\cite{Agrawal:2013uka, Debnath:2018azt} that using $\Delta_4$ as a kinematic observable helps break the degeneracy along the flat direction and determine the overall mass scale, and we therefore expect $\Delta_4$ to be a powerful variable when performing a neural network based analysis as well. Even so, we saw at the end of section~\ref{sec:HLanalysis} that the overall mass scale still has a large uncertainty when analyzing the data event-by-event, even with $\Delta_4$ as one of the HL variables. In this section we combine the power of $\Delta_4$ with an ensemble-based analysis to pin down the overall mass scale. 

We remind the reader that a histogram of $\Delta_4$ depends not only on the true value of $M_{\chi}$ in the data, but also the input value $M_{\chi}$ that is assumed when calculating $\Delta_4$. The core concept for the method we are about to present relies on training the neural network on the shape of this distribution as either the true or input value for $M_\chi$ is varied.

To determine the true value of $M_\chi$, we specify a range $[M_1, M_2]$ that we expect it to lie in, and we transform each event $x_i$ to a point in the following two-dimensional space, with $M_1$ and $M_2$ used as input values:  
\begin{equation}
    x_i \rightarrow (\Delta_4(x_i; M_1), \Delta_4(x_i; M_2)).
\end{equation}
Any data sample with the true value $M_\chi$ is then converted into a two-dimensional scatter plot, and with an appropriate binning, into a pixellated heat map. These heat maps are used to train the neural network, as the true $M_\chi$ is varied between $M_1$ and $M_2$, and $M_\chi$ being assigned as the output of the neural network during the training. In the testing phase, the output of the neural network is then taken as the measured value of $M_\chi$.

To minimize bias, we choose the relatively broad mass range of (100, 900) GeV for this last stage of our analysis. We use a step size of $\delta M_\chi = 20$ when varying $M_\chi$. We have checked that a smaller step size does not result in an improvement of the accuracy of the final result.

To generate the training data, we construct the heat maps described above from the data samples for each value of $m_\chi$ and $S/B$ of interest, after having applied the discovery cut (see section~\ref{sec:DLdeltam}). We restrict the heat maps to the range $-\Delta_{4,max} \leq \Delta_4 \leq \Delta_{4,max}$ for each sample. Events lying beyond this range are discarded. We refine the resolution of the heat maps until a saturation in performance is reached in terms of the loss function. In our case, this is achieved for a bin size of $\Delta_{4, max}/20$ along each of the two axes. Based on this number, we construct $41 \times 41$ pixel heat maps. The networks have 1681 nodes in the input layer, and three hidden layers with 100, 300, 100 nodes respectively.

\begin{figure}[h!]
     \begin{subfigure}[b]{0.45\textwidth}
         \centering
         \includegraphics[width=\textwidth]{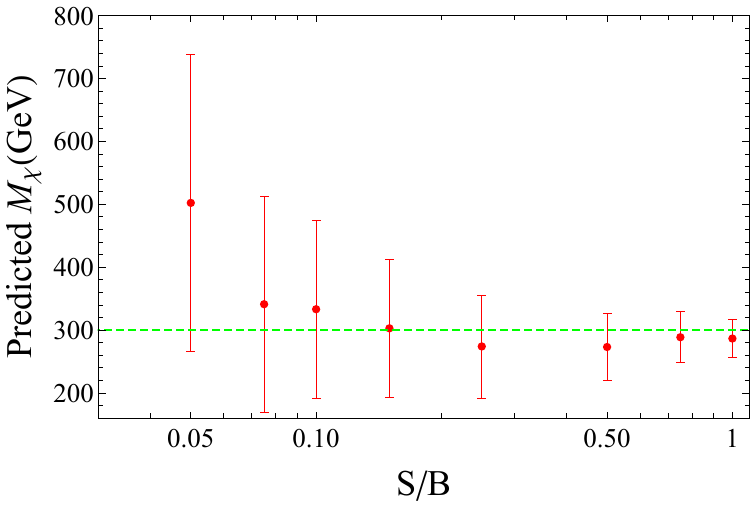}
         \caption{sDM}
         \label{fig: mass-sdm}
     \end{subfigure}
     \hfill
     \begin{subfigure}[b]{0.45\textwidth}
         \centering
         \includegraphics[width=\textwidth]{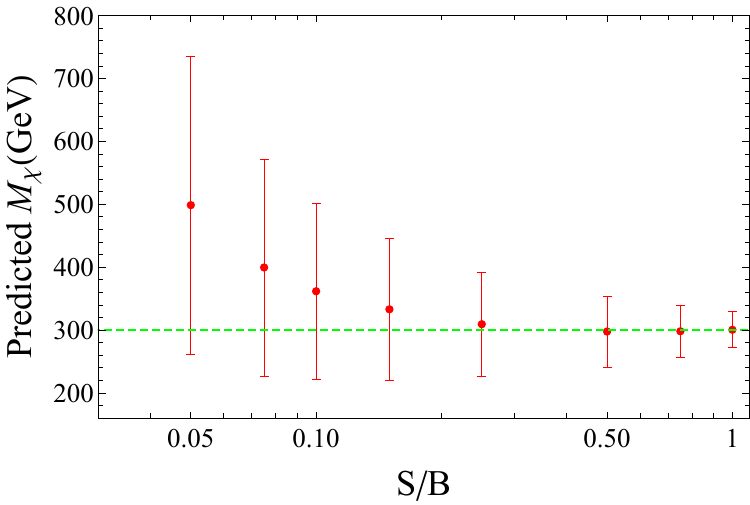}
         \caption{fDM}
         \label{fig: mass-fdm}
     \end{subfigure}
      
        \caption{The determined $M_\chi$ values and the uncertainties are plotted for $S/B \gsim 0.05$. Below this value, our methods cannot significantly narrow down the possible range of $M_\chi$ within our scan interval.}
        \label{fig: mass-error}
\end{figure}

Having completed the training phase, we move on to the determination of $M_\chi$ from the data. In the testing phase, the trained network is run on the data, with the estimated $S/B$ value obtained from equation~\ref{eq: cs-det} as input. In 1000 iterations for each value of $S/B$, the prediction for $M_\chi$ is obtained from the network output in each run. In figure~\ref{fig: mass-error}, we show the central value and standard deviation for $M_\chi$ obtained in this way over the $S/B$ range of interest. The central value is found to be within $1\sigma$ of the true value of $M_\chi$ over the entire $S/B$ range, with the precision approaching $20$~GeV as $S/B$ approaches 1.0.

\section{Conclusions}
\label{sec:conclusions}

We have applied machine learning techniques to optimize discovery sensitivity, spin determination and mass measurement in the SUSY-like decay chain of figure~\ref{fig:FeynmanDecay}. This event topology was chosen as a representative case for when signal cross sections are relatively low, and the final state particles do not have high $p_T$ due to a compressed signal spectrum. The decay chain is not long enough for algebraic reconstruction methods to be effective, and yet long enough that commonly used one-dimensional distributions of Lorentz-invariant or boost invariant kinematic observables do not capture the full amount of useful kinematic information.

In order to narrow down the parameter space, we started our analysis with a simple neural network that was effective in determining the mass gaps in the spectrum, and in enhancing signal over background for the second stage of the analysis. We have identified the kinematic observables that match the performance of a `black-box' network in the first stage of the analysis, confirming the importance of the observable $\Delta_4$ which had previously been proposed to be an effective observable for similar decay chains.

In the second, ensemble-based, stage of the analysis, we were able to achieve a much higher accuracy in determining the spins of the new particles compared to an event-by-event analysis. Similarly, we were also able to measure the overall mass scale, a significant challenge for methods based on commonly used  observables such as kinematic edges and endpoints, with an ensemble-based analysis. Once again, $\Delta_4$ proved to play a significant role in maximizing the precision of the measurement of the overall mass scale.

We point out to the reader that $\Delta_4$ is an $\mathcal{O}(8)$ polynomial of the Lorentz invariant pairs $m_{ij}$. To test the efficiency of $\Delta_4$ for discovery, a scan can be performed over polynomial functions of $m_{ij}$. Specifically, the binary cross-entropy loss can be minimized with respect to the coefficients of the polynomials of $m_{ij}$. We will address such global optimization in future work. 

The application of similar methods to more general event topologies and more challenging final state particles, to which traditional SUSY searches are not sensitive, is of great interest. Having provided a proof-of-concept with this analysis of an idealized final state, we will take on these challenges in future work.
\section*{Acknowledgements}

This paper is dedicated to the memory of Maaz Ul Haq. The authors are grateful to Konstantin Matchev for helpful discussions. The research of CK and RR is supported by the National Science Foundation Grant Numbers PHY-1914679 and PHY-2210562. The authors acknowledge the \href{http://www.tacc.utexas.edu}{Texas Advanced Computing Center} (TACC) at The University of Texas at Austin for providing HPC resources that have contributed to the research results reported within this paper.


\bibliographystyle{JHEP}
\bibliography{bib_PSML}{}
\end{document}